\documentclass[twocolumn]{aastex62}
\usepackage{graphicx}
\usepackage{amsmath}
\usepackage{url}
\usepackage{makecell}
\usepackage{threeparttable}
\usepackage{multirow}
\usepackage{xcolor}
\usepackage[utf8]{inputenc}
\usepackage{lipsum}
\newcommand{\yisx}{\color{red}}

\usepackage{ulem}
\newcommand{\GWT}{\texttt{GW-Universe Toolbox}}
\begin{document}
\title{A simulation study on the constraints of the Hubble constant using sub-threshold GW observation on double neutron star mergers}
\author{Yun-Fei Du}
\affiliation{Key Laboratory of Particle Astrophysics, Institute of High Energy Physics, Chinese Academy of Sciences, \\
19B Yuquan Road, Beijing 100049, People’s Republic of China}
\affiliation{University of Chinese Academy of Sciences, Chinese Academy of Sciences, Beijing 100049, People’s Republic of China}
\author{Shu-Xu Yi}
\affiliation{Key Laboratory of Particle Astrophysics, Institute of High Energy Physics, Chinese Academy of Sciences, \\
19B Yuquan Road, Beijing 100049, People’s Republic of China}
\author{Shuang-Nan Zhang}
\affiliation{Key Laboratory of Particle Astrophysics, Institute of High Energy Physics, Chinese Academy of Sciences, \\
19B Yuquan Road, Beijing 100049, People’s Republic of China}
\affiliation{University of Chinese Academy of Sciences, Chinese Academy of Sciences, Beijing 100049, People’s Republic of China}
\affiliation{National Astronomical Observatories, Chinese Academy of Sciences, Beijing 100012, People’s Republic of China}
\author{Shu Zhang}
\affiliation{Key Laboratory of Particle Astrophysics, Institute of High Energy Physics, Chinese Academy of Sciences, \\
19B Yuquan Road, Beijing 100049, People’s Republic of China}
\date{February 2023}
%\maketitle  
\begin{abstract}
     Gravitational waves observation with electromagnetic counterparts provides an approach to measure the Hubble constant which is also known as the bright siren method. Great hope has been put into this method to arbitrate the Hubble tension. In this study, we apply the simulation tool \GWT\, and modeling of the aLIGO-design background to simulate the bright siren catalogues of sub-threshold double neutron star mergers with potential contamination from noise and dis-pairing between gravitational waves and electromagnetic counterparts. The Hubble constant and other cosmology parameters are thus inferred from the simulated catalogues with a Bayesian method. From our simulation study, we reach the following conclusions: 1) the measurement error of the $H_0$ decreases with a lower signal-to-noise ratio threshold (or equivalently the $P_{\rm astro}$) in the region where $P_{\rm astro} \gtrsim $ 0.1, while the inferred most probable $H_0$ trends to bias towards larger values; and 2) other higher order cosmological parameters such as $\Omega_{m}$ remain unconstrained even with the sub-threshold catalogues.
     %The results are the $\Omega_{m}$ or other cosmology parameters could not be constrained, the measurement error of the Hubble constant decreases with the increase of the signal-to-noise ratio threshold ($P_{\rm astro} \gtrsim $ 0.7), the bias of the measurement increases with the increase of the signal-to-noise ratio threshold. The Hubble constant posterior is greater than the theory value,  that is due to the upper limit of distance measurement. 
     We also discuss adding the network of the gravitational wave detectors to the simulation tool and the electromagnetic counterparts follow-up efficiency simulation, which will improve our work in the future. 
\end{abstract}

%\keywords{Gravitational waves(678) --- Hubble constant(758) --- Cosmological parameters(339) --- LIGO(920)}

\section{Introduction}

The expansion rate of the Universe at present, known as the Hubble constant ($H_0$), is one of the most basic cosmological parameters. The determination of $H_0$ is one of the core tasks of observational cosmology, which involves independent estimation of red-shift and distance. While the former is usually obtained from optical measurement of spectral lines in a rather straight forward way, a reliable independent measurement of distance is quite challenging. One well-established measurement method is the cosmic distance ladder employing Cepheid variables, red-giant stars and type Ia supernovae \citep[e.g.][]{1993ApJ...413L.105P, 1995ApJ...438L..17R, 2003ApJ...590..944W, 2016ApJ...826...56R, 2017arXiv170201118J, 2022ApJ...934L...7R}. Based on this method, the best constraints on $H_0$ coming from the SH0ES Cepheid-SN distance ladder is $73.04\pm 1.04$\,\rm{km\,s$^{-1}$\,Mpc$^{-1}$} \citep[]{2022ApJ...934L...7R}. Another completely different approach to $H_0$ is to model the anisotropy of the Cosmic Microwave Background (CMB) with the $\Lambda$ cold dark matter ($\Lambda$-CDM) model \citep[e.g.][]{2007ApJS..170..377S, 2013ApJS..208...19H, 2014A&A...571A..16P, 2016A&A...594A..13P, 2020A&A...641A...6P}, with $H_0$ one of the model parameters. The latest result of $H_0$ from this approach is $67.4\pm \,0.5$\,\rm{km\,s$^{-1}$\,Mpc$^{-1}$} \citep[]{2020A&A...641A...6P}. The obvious discrepancy between the $H_0$ determined by the above mentioned two methods (at more than $\sim$ 3$\sigma$ confidence level) implies that, there are either overlooked systematic in these methods, or our current modeling of the Universe is incomplete \citep[see][for a recent review]{2019NatAs...3..891V}. 

The gravitational waves (GW) observation provides an alternative approach to the measurement of $H_0$, which will cast light on the Hubble constant tension. The distance of the GW source can be determined with the so-called standard siren method \citep[]{1986Natur.323..310S, 2005ApJ...629...15H}: The strain amplitude $h_{+}$ and $h_{\times}$ of the GW can be written as: 
\begin{equation}
h_{+}(\tau_{\rm obs})=h_c(\tau_{\rm obs})\,\frac{1+\cos^2\iota}{2}\,\cos[\Phi(\tau_{\rm obs})],
\label{hplus}
\end{equation}
\begin{equation}
h_{\times}(\tau_{\rm obs})=h_c(\tau_{\rm obs})\,\cos\iota \,\sin[\Phi(\tau_{\rm obs})],
\label{htimes}
\end{equation}
where 
\begin{equation}
h_c(\tau_{\rm obs})=\frac{4}{D_{\rm{L}}}(\frac{G\mathcal{M}_{c,z}}{c^2})^{5/3}(\frac{\pi f_{\rm gw}^{\rm obs}(\tau_{\rm obs})}{c})^{2/3},
\label{hc}
\end{equation}
here $\tau_{\rm obs}$ is the time in the observer reference frame, $D_{\rm{L}}(z)$ is the luminosity distance to the GW source, $G$ is the gravitational constant, $\iota$ is the orbital inclination of binary stars, $\Phi$ is the phase of the GW signal,  $\mathcal{M}_{c,z}=(1+z)(m_1 m_2)^{3/5}(m_1 + m_2)^{-1/5}$ is the red-shift “chirp” mass and $f_{\rm gw}^{\rm obs}$ is the frequency of GW in the observer frame. The $\mathcal{M}_c(z)$, $\iota$, $f_{\rm gw}^{\rm obs}$ and  $\Phi$ can be inferred by matching modeled waveform to the GW strain data. The characteristic strain $h_c$ is inversely proportional to the luminosity distance $D_{\rm{L}}$, therefore, {\yisx} can be inferred from the GW data fitting. The above method of $D_{\rm L}$ inference is known as the standard siren method. The red-shift of the source can be, on the other hand, obtained from the electromagnetic (EM) counterparts of the GW or its host galaxy (“bright sirens” \citep[e.g.][]{2010ApJ...725..496N, 2021PhRvL.126q1102F, 2019PhRvD.100j3523M}) or otherwise statistically based on galaxy clustering (“dark sirens” \citep[e.g.][]{2008PhRvD..77d3512M, 2019ApJ...871L..13F, 2022arXiv220311756G}). In this paper, we will focus on the bright sirens, i.e., GWs with individually identified EM counterparts (EMC). 

DNS mergers generate GW, can also be observed as prompt gamma-ray bursts (GRBs, usually with short duration, but some with longer duration are also believed to be DNS merger-induced, e.g., GRB 211211A \citep[]{2022arXiv220502186X}), together with their afterglow in X-ray, optical and radio bands. A kilonova \citep[]{2010MNRAS.406.2650M} is another kind of optical counterpart expected from DNS mergers. Those sources in EM wave bands are refereed to as EM counterparts (EMCs) of the GW events. For instance, the GRB170817A was found 1.7 s after the GW170817, the kilonova AT 2017gfo was found hours after the GW and lasted for about 10 days until it could not be seen \citep[]{2017ApJ...848L..12A}. The detection of the prompt GRB provides extra early localization and is very helpful for the host galaxy identification, wherein an independent red-shift measurement can be obtained. However, due to their highly beamed emission, the coincidence fraction between a GW and a GRB is expected to be low \citep[]{2022arXiv220814156H}. On the other hand, those EMCs in afterglows and kilonovae are expected to be more accessible for most of the GW signals, due to their nearly isotropic radiation distribution. For instance, a kilonova is expected to be detected as far as a few hundreds Mpc \citep[]{2021MNRAS.504.1294S, 2022ApJ...927...50R, 2022ApJ...927..163C}, which overlaps with the detection range of DNS with the 2nd generation GW detectors. Therefore, kilonovae are used as the EMCs of GW events in our work.

GW170817 is the only GW event which has an confirmed EMC. This GW with EMC is generally believed to originate from the merger of a system of double neutron stars (DNS). Based on the bright siren with this single event, $H_0$ is constrained to be 70.0$^{+12.0}_{-8.0}$\,\rm{km\,s$^{-1}$\,Mpc$^{-1}$} at the 1-$\sigma$ level \citep[]{2017Natur.551...85A}; the large uncertainty make the result compatible with both values from cosmic distance ladder and CMB methods. While simulations \citep[e.g.][]{2018Natur.562..545C, 2021PhRvL.126q1102F} found that with more events of GW with EMC, bright sirens can give discriminatory $H_0$ with small enough uncertainties. \cite{2018Natur.562..545C} has simulated the LIGO/Virgo, KAGRA and LIGO-India's GW observation and found that the uncertainty of $H_0$ will scale roughly as 13$\%$/$\sqrt{N}$, where $N$ is the number of DNS bright sirens. \cite{2021PhRvL.126q1102F} showed that $\sim$ 50 DNS bright sirens by LIGO and Virgo can independently arbitrate the Hubble tension. \cite{2012PhRvD..86d3011D} performed a simulation of the advanced world-wide network of GW observations. Their simulation showed that a few tens of bright sirens will constrain the $H_0$ to an accuracy of $\sim$ 4-5$\%$ at 95$\%$ confidence. However, the energy density parameters $\Omega_m$ and $\Omega_{\Lambda}$ will not be constrained by the upcoming network of GW observatories, regardless of the number of events of GW with EMC.

\cite{2022arXiv220814156H} showed that decreasing the GW detection threshold to a marginally tolerable purity will double the rate of joint detection of GRB and GW. It is also intuitive to suspect that the same sub-threshold strategy will lead to a larger multi-messenger catalogue with GW and their EMC. An obvious trade-off of lowering the signal-to-noise ratio (S/N) threshold is that more possible fake signals from the background noise will be included as GW candidates, which is often referred to as the decrease of purity of the catalogue, where the purity is defined as the number of GW candidates that are genuinely astronomical divided by the total number of candidates in the catalogue. For our specific purpose of bright siren $H_0$ constraint, the larger sample of GW and EMC will result a more precise constraint, while the less purity of catalogue may in turn hinder the very purpose. We therefore find it intriguing to study which factor prevails in a certain range of S/N thresholds, and such a study will serve as a practical strategy to maximise the science outputs with current GW/EMC multi-messengers observation capabilities. %Sub-threshold is one useful method to increase the number of BNS we detected. When we observe GW signals, we will set a signal-to-noise ratio (S/N) threshold to remove the false signals which may come from noise. The false-alarm-rate (FAR) is the expected rate of events per time due to noise which would be the fake signal mistaken for the GW events candidate\citep[]{2020CQGra..37e5002A}. It is a exponential decay function of S/N threshold. Therefore, if we lower the S/N threshold, we will observe more GW signals, at the same time, a lot of false signals are introduced. This is a trade-off between larger BNS catalogues and more potential contamination. When we use the sub-threshold catalogues to constrain the $H_0$, the uncertainties of $H_0$ will decrease and the deviation of $H_0$ may increase. We want to find the optimised S/N threshold when we use the sub-threshold catalogue to constrain the H0 by standard sirens. This is an important reference when we do the sub-threshold GW signal search. 

Based on the above-mentioned motivation, we perform the study with simulation. We generate synthetic GW catalogues with different S/N thresholds using a software package \GWT. In order to simulate the effects of fake signal contamination, we incorporate a new GW population with \GWT, whose S/N distribution resembles that of the aLIGO-design backgrounds signals in O1-O2 runs. The obtained synthetic GW catalogues are mixed with candidates from both the astronomical population and the noise population. For a fraction of GW from the astronomical population, we assume they are associated with observable EMCs. Each of the EMCs is re-assigned to a GW candidate from the whole catalogue, based on their estimated localization and temporal coincidence. It is a representation of the EMC identification process in a real multi-messenger follow-up campaign, where mis-pairing between GW and EMC is likely.

In Section 2, we introduce the simulation tool, modeling of the aLIGO-design background and the method to simulate sub-threshold DNS bright siren catalogues with potential contamination from noise and dis-pairing between GW and EMC. In Section 3, we make use of the bright siren catalogues to perform Bayesian inference on $H_0$ and other cosmology parameters. Section 4 focuses on the results, i.e., the sub-threshold catalogues' constraint on $H_0$ with different sub-threshold observation set-ups. Then we give our general conclusion on the prospect of the sub-threshold bright siren on $H_0$ measurement. In Section 5, we discuss the aspects that can be improved in the future, and summarize the findings of the paper. 

\section{Simulating sub-threshold DNS catalogues contaminated with background}
\subsection{The gravitational wave universe toolbox}
\begin{figure}
    \centering
    \includegraphics[width=.5\textwidth]{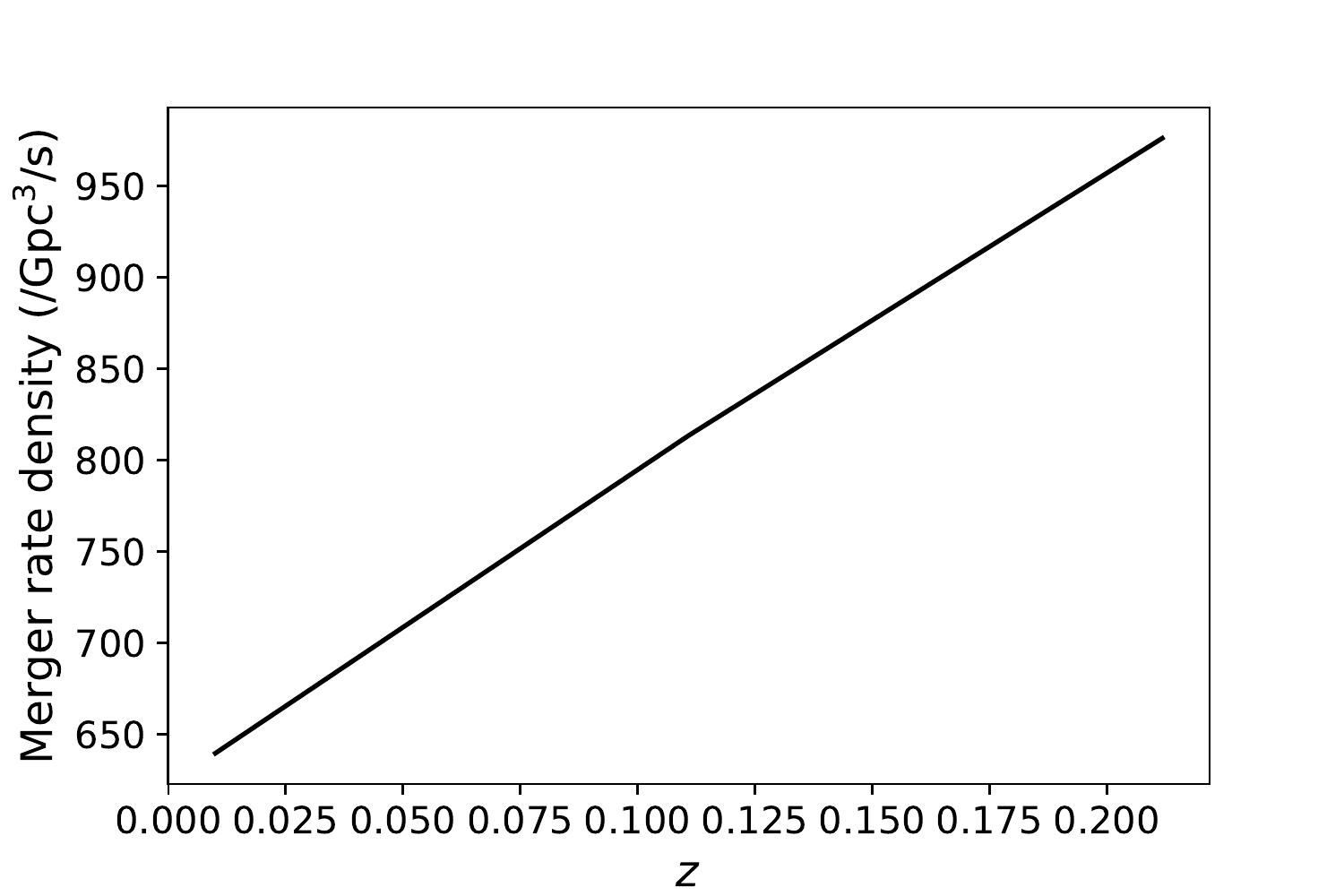}
    \caption{The cosmic merger data density as function of red-shift in DNS-Pop I model.}
    \label{fig:my_label}
\end{figure}

\GWT\, is a comprehensive software package that simulates the observation on common GW sources from nHz to kHz, with various of detecting methods. We apply its ground-based detector module, where we can choose the GW detectors from a default collection of aLIGO-O3, aLIGO-design, advanced Virgo-design, KAGRA-design, Einstein Telescope and Cosmic Explorer. The targeted GW population can be chosen from binary black hole (BBH) mergers, DNS and black hole-NS mergers. With a user specified S/N threshold and observation duration, \GWT\, will return a catalogue of simulated GW events with their physical and geometrical parameters. The uncertainties of the parameters can be also estimated based on a Fisher information matrix (FIM) based algorithm. This module is designed to be also flexible, so that users can specify their own noise curves of detectors, as well as the source population models \citep[for details see][]{2022A&A...663A.155Y} For the purpose of our study, we simulate the observation with aLIGO-design, and astronomically originated GW sources from the default DNS population model (DNS-Pop I). The cosmic merger data density as function of red-shift in the DNS-Pop I model is plotted in figure \ref{fig:my_label}.  

\subsection{modelling the background as a source population}
If we lower the S/N threshold $\rho_\star$, more noise from the background of the detector will be included as signals. Thus the false-alarm rate (FAR) increases with lower $\rho_\star$. \cite{2018ApJ...861L..24L} found the FAR for LIGO Handford and LIGO Livingston in O1-O2 runs as function of $\rho_\star$ can be fitted with an exponential function:
\begin{equation}
\rm{FAR}=\rm{FAR}_8\times\exp\big[-\frac{\rho_\star-8}{\alpha}\big],
\label{LynchFAR}
\end{equation}
where FAR$_8$ is the FAR when $\rho_\star=8$. Equation \ref{LynchFAR} is equivalent to that the accumulative frequency of the background ``event" population is:
\begin{equation}
    \mathcal{C}(\rho)\equiv\frac{1}{T}\int^\infty_\rho f(\tilde{\rho})d\tilde{\rho}=\rm{FAR}_8\times\exp\big[-\frac{\rho-8}{\alpha}\big],
    \label{eq:accum}
\end{equation}
where $f(\rho)$ is the differential number distribution of $\rho$ in a given observation duration $T$, which can be obtained by taking derivative on both sides of Equation \ref{eq:accum}:
\begin{equation}
    f(\rho)/T=\frac{\rm{FAR}_8}{\alpha}\times\exp\big[-\frac{\rho-8}{\alpha}\big].
    \label{eq:frequency}
\end{equation}

We want to construct a GW source population, whose accumulate merger rate distribution as function of $\rho_\star$ mimics that in Equation \ref{eq:accum}. Assume the number of events from that fake population is $\mathcal{N}$ over $T$, we thus have:
\begin{equation}
    d\mathcal{N}=f(\rho)d\rho.
\end{equation}
Assuming $\mathcal{N}$ is function of the binary's masses and red-shift, we therefore have:
\begin{equation}
    d\mathcal{N}=\frac{\partial\mathcal{N}}{\partial m_1}dm_1+\frac{\partial\mathcal{N}}{\partial m_2}dm_2+\frac{\partial\mathcal{N}}{\partial z}dz=f(\rho)d\rho.
    \label{eq:partial}
\end{equation}

Since for DNS the mass range is small, we assume that the $\mathcal{N}$ depends weakly on masses and attribute all the $\rho$ dependence of $\mathcal{N}$ on the luminosity distance, or equivalently, on $z$. As a result, equation \ref{eq:partial} becomes:
\begin{equation}
    \frac{\partial\mathcal{N}}{\partial z}=f(\rho)\frac{d\rho}{dz}.
    \label{eq:linkNtorho}
\end{equation}
Assume that the differential distribution of sources at the observer is:
\begin{equation}
    d\mathcal{N}(m_1, m_2, z)=T\times d\dot{n}(m_1, m_2, z),
    \label{eq:7}
\end{equation}
and suppose the differential rate d$\dot{n}$ depends on masses and $z$ separately, we obtain
\begin{equation}
    d\dot{n}(m_1,m_2,z)=p(m_1,m_2)d\dot{n}(z).
\end{equation}

Combining Equations (\ref{eq:frequency}, \ref{eq:linkNtorho} and \ref{eq:7}), we have
\begin{equation}
    p(m_1,m_2)\frac{d\dot{n}(z)}{dz}=\frac{\rm{FAR}_8}{\alpha}\times\exp\big[-\frac{\rho-8}{\alpha}\big]\frac{d\rho}{dz}.
    \label{eq:plinkfar}
\end{equation}
Integrate over mass range on both sides of Equation \ref{eq:plinkfar}, we have
\begin{equation}
    \frac{d\dot{n}(z)}{dz}=\big(m_{\rm{high}}-m_{\rm{low}}\big)^2\frac{\rm{FAR}_8}{\alpha}\times\exp\big[-\frac{\rho-8}{\alpha}\big]\frac{d\rho}{dz}.
    \label{eq:10}
\end{equation}
We know that $\rho$ is inversely proportional to the luminosity distance $D$. Therefore $\rho$ can be represented as:
\begin{equation}
    \rho\approx8\frac{D_{\rm{ref}}}{D},
    \label{eq:11}
\end{equation}
where $D_{\rm {ref}}$ is the reference luminosity distance where the $\rho$ of a DNS is approximately 8. Take the above relationship into Equation \ref{eq:10}, we have:
\begin{equation}
    \frac{d\dot{n}(z)}{dz}=\big(m_{\rm{high}}-m_{\rm{low}}\big)^2\frac{\rm{FAR}_8}{\alpha}\times\exp\big[-\beta\big(\frac{D_{\rm{ref}}}{D}-1\big)\big]\big|\frac{d\rho}{dD}\big|\frac{dD}{dz}.
    \label{eq:long}
\end{equation}
In the above equation, $\beta=8/\alpha$. Work out $d\rho/dD$ using Equation \ref{eq:11} and denote
\begin{equation}
    8\big(m_{\rm{high}}-m_{\rm{low}}\big)^2\frac{\rm{FAR}_8}{\alpha}=\mathcal{R}_n,
\end{equation}
we rewrite Equation \ref{eq:long} as:
\begin{equation}
    \frac{d\dot{n}(z)}{dz}=\mathcal{R}_n\exp\big[-\beta\big(D_{\rm{ref}}/D(z)-1\big)\big]\frac{D_{\rm{ref}}}{D(z)^2}\frac{dD(z)}{dz},
    \label{eq:long2}
\end{equation}
where the luminosity distance is denoted as function of the red-shift. 

Denote the population merger rate density as function of $z$ as $\mathcal{R}(z)$. The relation between $\mathcal{R}$ and $\frac{d\dot{n}}{dz}$ is:
\begin{equation}
    \frac{d\dot{n}(z)}{dz}=\mathcal{R}(z)\frac{1}{1+z}\frac{dV_{\rm{c}}}{dz},
    \label{eq:15}
\end{equation}
where $dV_{\rm{c}}$ is the differential comoving volume of the universe. Combining Equations (\ref{eq:long2}) and (\ref{eq:15}) we have the cosmic merger rate density:
\begin{equation}
    \mathcal{R}(z)=\mathcal{R}_n\exp\big[-\beta\big(D_{\rm{ref}}/D(z)-1\big)\big]\frac{D_{\rm{ref}}}{D(z)^2}\frac{dD(z)}{dz}\frac{1+z}{dV_{\rm{c}}/dz}.
    \label{eq:final}
\end{equation}

\begin{figure}
    \centering
    \includegraphics[width=.5\textwidth]{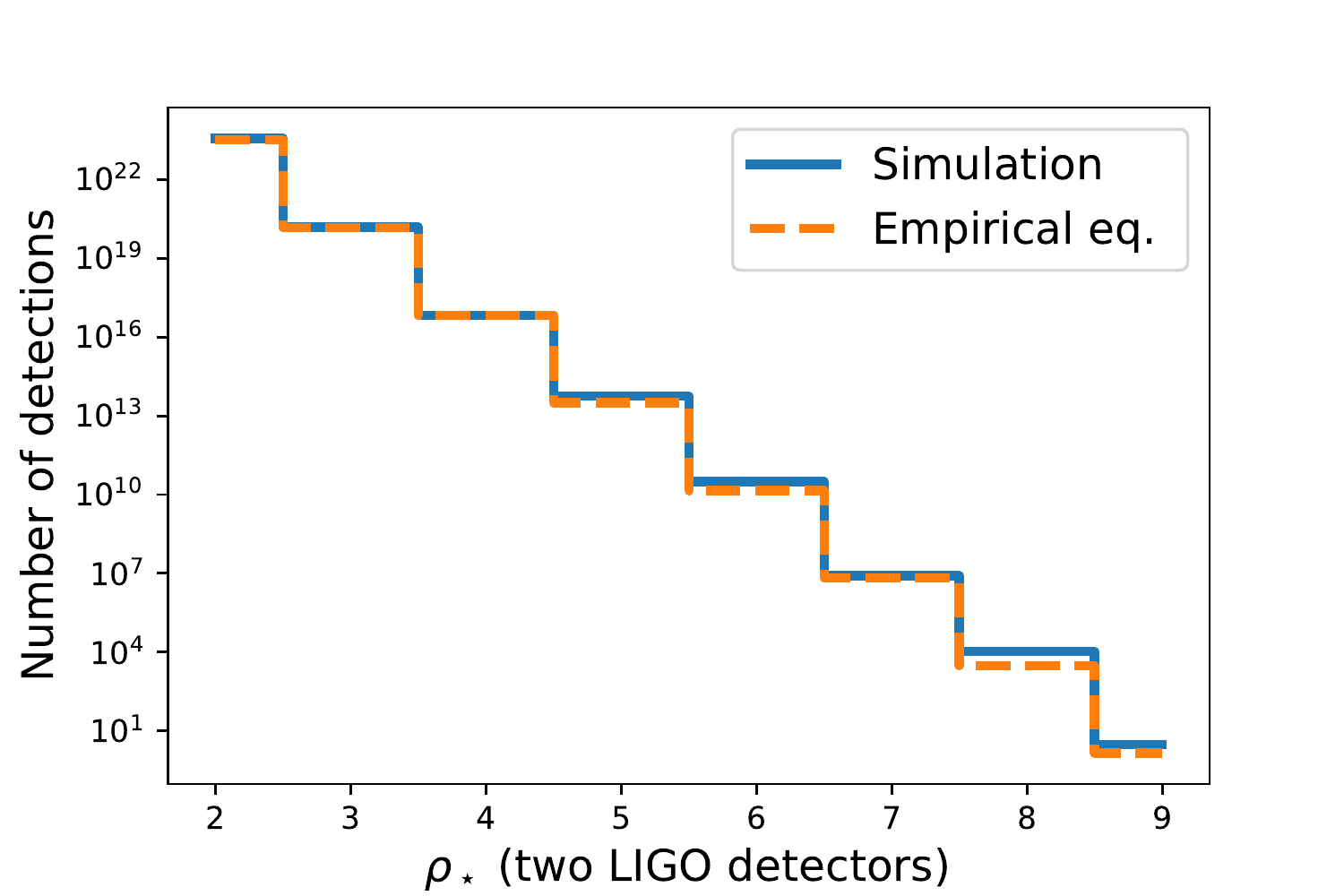}
    \caption{The background DNS population model simulation (blue solid line) in this study and the FAR in Equation \ref{LynchFAR} (red dashed line) obtains the number of the detection by two LIGO detectors as a function of $\rho_{\star}$, respectively.}
    \label{fig:accum}
\end{figure}

So far, we have obtained the cosmic merger rate of the equivalent DNS population originating from the background noise. We further assume the mass dependence of $p(m_1, m_2)$ is uniform over the range $(m_{\rm{low}}, m_{\rm{high}})$. \cite{2018ApJ...861L..24L} found that for DNS, FAR$_8=3\times10^4$\,yr$^{-1}$, $\alpha=0.13$. This provides clues for our choices of values of $\beta$ and $\mathcal{R}_n$, where we use $\mathcal{R}_n=1\times10^6$\,yr$^-1$, $\beta=61.5$, $D_{\rm{ref}}=1.34\,$Mpc, and $m_{\rm{low}}=0.5\,M_\odot$, $m_{\rm{high}}=3\,M_\odot$. We plug the above population model into the \GWT, and simulate the detection of such a population of DNS with LIGO. Note that in Equation \ref{LynchFAR}, $\rho$ is for two LIGO detectors' combined values, but \GWT\, simulates a single detector. So we multiply a factor $\sqrt{2}$ to the single detector $\rho$ to convert it to the combined one. The number of detection as a function of $\rho_\star$ is plotted in Figure \ref{fig:accum}.

It can be seen from Figure \ref{fig:accum} that our fake DNS population model can represent the FAR in LIGO well. Although our background population model is calibrated against the FAR function found in O1-O2, we will still employ it in this simulation study for the designed aLIGO.
\begin{figure}
    \centering
    \includegraphics[width=.5\textwidth]{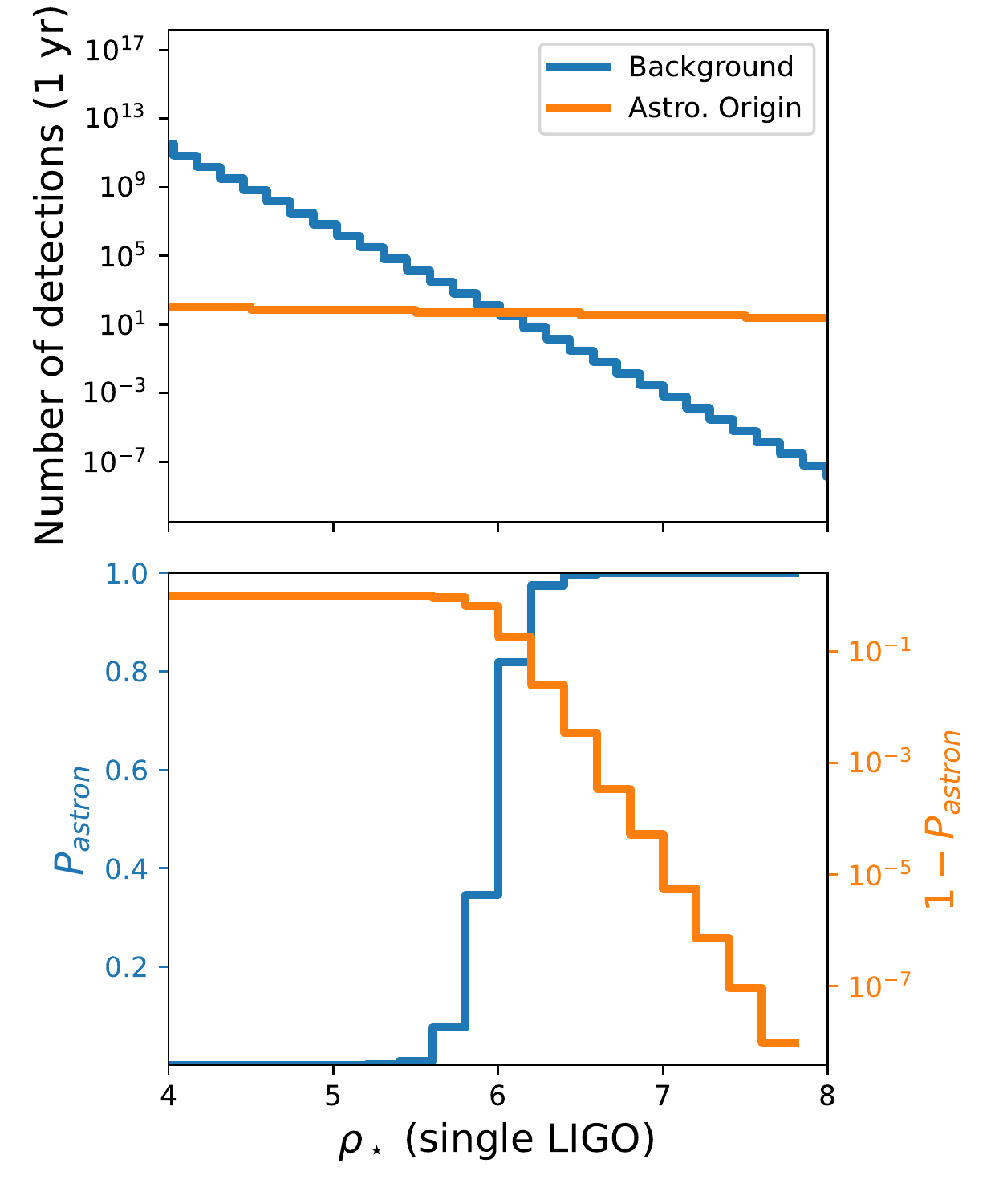}
    \caption{The top panel is the number of detection from the background DNS population model (solid blue line) and the astronomical origin model (``Pop-A", solid orange line) in one year versus $\rho_{\star}$ for a single LIGO. The bottom panel is $P_{\rm astro}$ (solid blue line) and $1-P_{\rm astro}$ (solid orange line) calculated with Equation \ref{eq:Pastro} versus the $\rho_{\star}$ for a single LIGO.}
    \label{fig:2figures}
\end{figure}

In order to compare the astronomically originated population with the background, we use the \GWT\, to simulate a population of DNS with the default population model ``Pop-A", and the parameters for the population model are: $R_n=300\,$Gpc$^{-3}$/yr, $\tau=3$\,Gyr, $m_{\rm{mean}}=1.4\,M_\odot$, $m_{\rm{scale}}=0.5\,M_\odot$, $m_{\rm{low}}=1.1\,M_\odot$, $m_{\rm{high}}=2.5\,M_\odot$, $\sigma_{\rm{\chi}}=0.1$. See \cite{2021arXiv210613662Y} for details for the population model \footnote{or find the description on this web-page: \url{https://gw-universe.org/population_model.html}.}. In Figure \ref{fig:2figures}, the upper panel shows the comparison between the numbers of sources from astronomical origin and from the background, as function of the $\rho_\star$; in the lower panel, we show the purity ($P_{\rm{astro}}$) as calculated with:
\begin{equation}
    P_{\rm{astro}}(\rho_\star)=\frac{N_{\rm {astro}}(\rho_\star)}{N_{\rm{astro}}(\rho_\star)+N_{\rm{background}}(\rho_\star)}.
    \label{eq:Pastro}
\end{equation}
%Since the $N_{\rm{background}}(\rho_\star)$ increases rapidly with the $\rho_\star$ decreases, we
We define $P^\prime_{\rm{astro}}$ as the differential purity of the marginal candidates in the neighborhood of $\rho_\star$:
\begin{equation}
    P^\prime_{\rm{astro}}(\rho_\star)=\frac{\Delta N_{\rm{astro}}(\rho_\star)}{\Delta N_{\rm{astro}}(\rho_\star)+\Delta N_{\rm{background}}(\rho_\star)}.
    \label{eq:Pprime}
\end{equation}

From Figure \ref{fig:2figures}, we find that at $\rho_\star<6$, the background events fast out-number the astronomical ones, and the purity of the detected catalogue becomes low. 

\subsection{The synthetic bright siren catalogues}

%From figure \ref{fig:2figures}, we find the minimum S/N threshold can be 6, where the number of background sources is comparable to the number of astronomical sources, because even smaller will introduce a lot of background sources. Therefore, we choice 6 as the S/N threshold to produce the initially GW events by \GWT\,\, and aLIGO-design background model. ({\yisx 6 is the lowest S/N in our simulations. Here we are only introducing the noise population, and not doing the simulation yet, so we can move this paragraph to later section.}) 
%modeling the misidentification of the EMC of GW

\begin{figure}
    \centering
    \includegraphics[width=10cm]{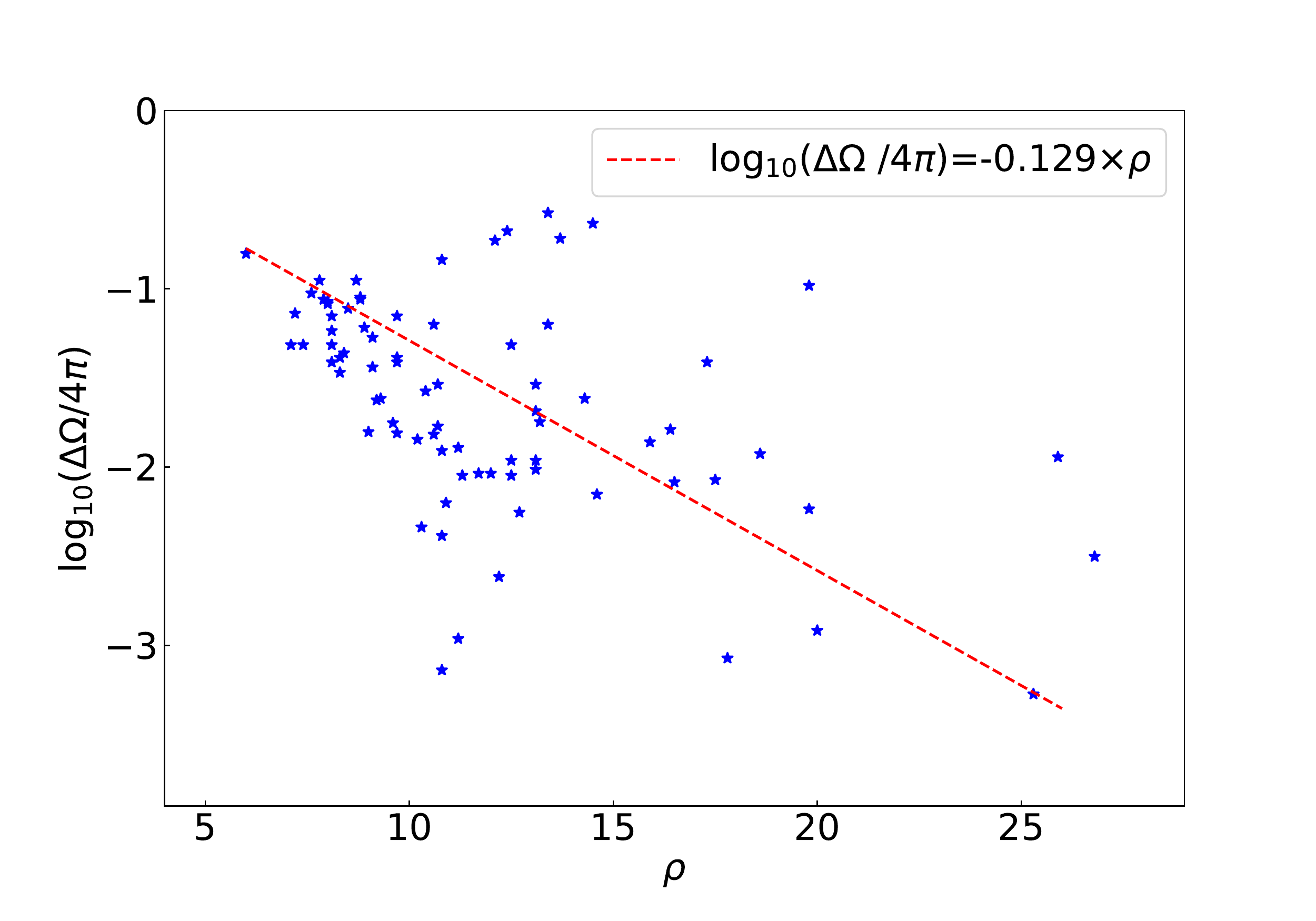}
    \caption{The relationship between the sky area angle and the $\rho$ of the GW observed in the LIGO/Virgo O3 period. The square of the correlation coefficient of the linearly fit is 0.883.}
    \label{fig:SNR empirical formula}
\end{figure}

A typical GW sky localization area is a few hundreds square degrees \citep[]{2021arXiv211103606T}, and the coincident time window for a GW chirp and a typical EMC can be as large as $\sim$ week \cite[]{2017ApJ...848L..12A}. Therefore mis-pairing between a EMC and a GW candidate is possible in the real follow-up observation campaign, when the catalogues in both bands are large. In order to include the potential mis-pairing between EMC and the GW candidates, we perform the following steps: 
\begin{enumerate}
    \item Simulating a catalogue of GW candidates, which is composed of astronomical origins and background population. For each GW candidate, randomly assign their sky coordinates ($\theta_i$, $\varphi_i$) uniformly on the celestial sphere, and their detected time ($t_i$) uniformly in the observation duration. 
    \item From the simulated astrophysical origin GW, we randomly assign a fraction of $\eta_{\rm{EMC}}$ to possess detectable EMC; the redshift of each EMC is calculated based on an underlying cosmological model (Plank18 \citep[]{2020A&A...641A...6P} in our practice) given the simulated luminosity distance of the corresponding GW candidate. 
    \item Up to this point, each EMC is associated with its true GW counterpart. Now we want to include the possibility of mis-pairing as in reality. For the $i$-th EMC, we find a pool of potential GW counterparts $\{G_j\}_i$ out of the overall GW catalogue. The GW pool consists of GWs which have their $\delta\Omega_{ij}<\Delta\Omega_i$ and $\delta t_{ij}<\Delta T_{\rm EMC}$, where $\delta\Omega_{ij}$ is the angular separation between the $i$-th GW candidate and $j$-th EMC, and $\Delta\Omega_i$ is the uncertainty of the sky location of the $i$-th GW; $\delta t_{ij}$ is the time lag between $j$-th EMC and the $i$-th GW candidate, and $\Delta T_{\rm EMC}$ is the coincident time window between a GW and its possible EMC. 
    
    \item From this pool, we randomly find one with a weight of $P^\prime_{\rm astro}$, and assign it as the counterpart of the $j$-th EMC.
%     \begin{equation}
%   C\equiv \frac{\Delta \Omega}{4\pi}\times\frac{\Delta T_{EMC}}{T_{obs}}\times\frac{\Delta D}{D},
%   \label{eq:22}
% \end{equation}
% where $\Delta\Omega$ is the sky-localization of the GW,  $\Delta T_{\rm EMC}$ is the coincident time window between a GW and its EMC. When the time lag between a EMC and a GW is larger than $\Delta T_{\rm EMC}$, they will not be identified as the counterparts of each other; 
%     \item 
\end{enumerate}
%We define the coincidence rate ($C$) to describe the probability that a GW is misidentified. We think the coincidence rate is the function of the sky area angle of the GW, the time resolved of the EMC and the distance resolved of the GW:
%\begin{equation}
%   C\equiv \frac{\Delta \Omega}{4\pi}\times\frac{\Delta            %T_{EMC}}{T_{obs}}\times\frac{\Delta D}{D},
%\end{equation}
%\begin{equation}
%  C=\frac{\Delta \Omega}{4\pi}*\frac{2\,w}{1\,yr}*1
%  \label{eq:22}
%\end{equation}
%where $\Delta T_{EMC}$ is the duration of the kilonova of the GW, which take two weeks as the typical duration. $\frac{\Delta \Omega}{4\pi}$ is the angular resolution of the GW. $T_{obs}$ is the observation time, in this paper, it takes 2 or 5 years. The distance positioning accuracy of the GW is too poor to $\Delta D /D $ $\sim$ 1. In equation \ref{eq:22}, for get the $C$, we still need know how to calculate $\Delta \Omega$. ({\duyf More explain why the $\Delta \Omega$ can be replaced by S/N} 
The uncertainties of sky-localization $\Delta\Omega$ are estimated by \GWT\, using a FIM based method. However, this estimation is based on single detector. While in reality, the localization can be much better using a network of detectors with either arriving time triangulation \citep{2011CQGra..28j5021F} or phase based method \citep{2020LRR....23....3A}. For this reason, we will not use the $\Delta\Omega$ given by the \GWT, but to interpolate the empirical relationships between the $\rho$ and $\Delta\Omega$ in the real observed catalogue of GWTC-3 \citep[]{2021arXiv211103606T}, using the $\rho$ of each GW returned from \GWT. By analyzing the relationship between the sky area angle and the $\rho$ of the GW observed in the LIGO/Virgo O3 period, we found an empirical formula:
\begin{equation}
   \frac{\Delta \Omega}{4 \pi}=10^{-0.129 \times \rho}.
   \label{eq:23}
\end{equation}
Thus, $\Delta \Omega$ is calculated by $\rho$ through this formula. The relationships between the $\rho$ and $\Delta\Omega$ in the real observed catalogue of GWTC-3 is shown in Figure \ref{fig:SNR empirical formula}.

Here we use a factor $\eta_{\rm EMC}$ to represent the ratio between the numbers of the EMC which can be identified in the follow-up observation, and that of the total GW candidates. This $\eta_{\rm EMC}$, which we also refer to as the EMC follow-up efficiency, is difficult to evaluate, given its sophisticate links to EMC models, telescopes performance and multi-wavelength observation campaign (see discussion in section 5.2). In this study, we let $\eta_{\rm EMC}$ be a free parameter, which we take two fiducial values $50\%$ and $100\%$. The former one roughly represent the $\eta_{\rm EMC}$ found in the current DNS catalogue so far (GW170817 and GW190425 \citep{2020ApJ...892L...3A}), and the latter represents the extreme case with an optimal follow-up efficiency.

\begin{figure}
    \centering
    \includegraphics[width=.5\textwidth]{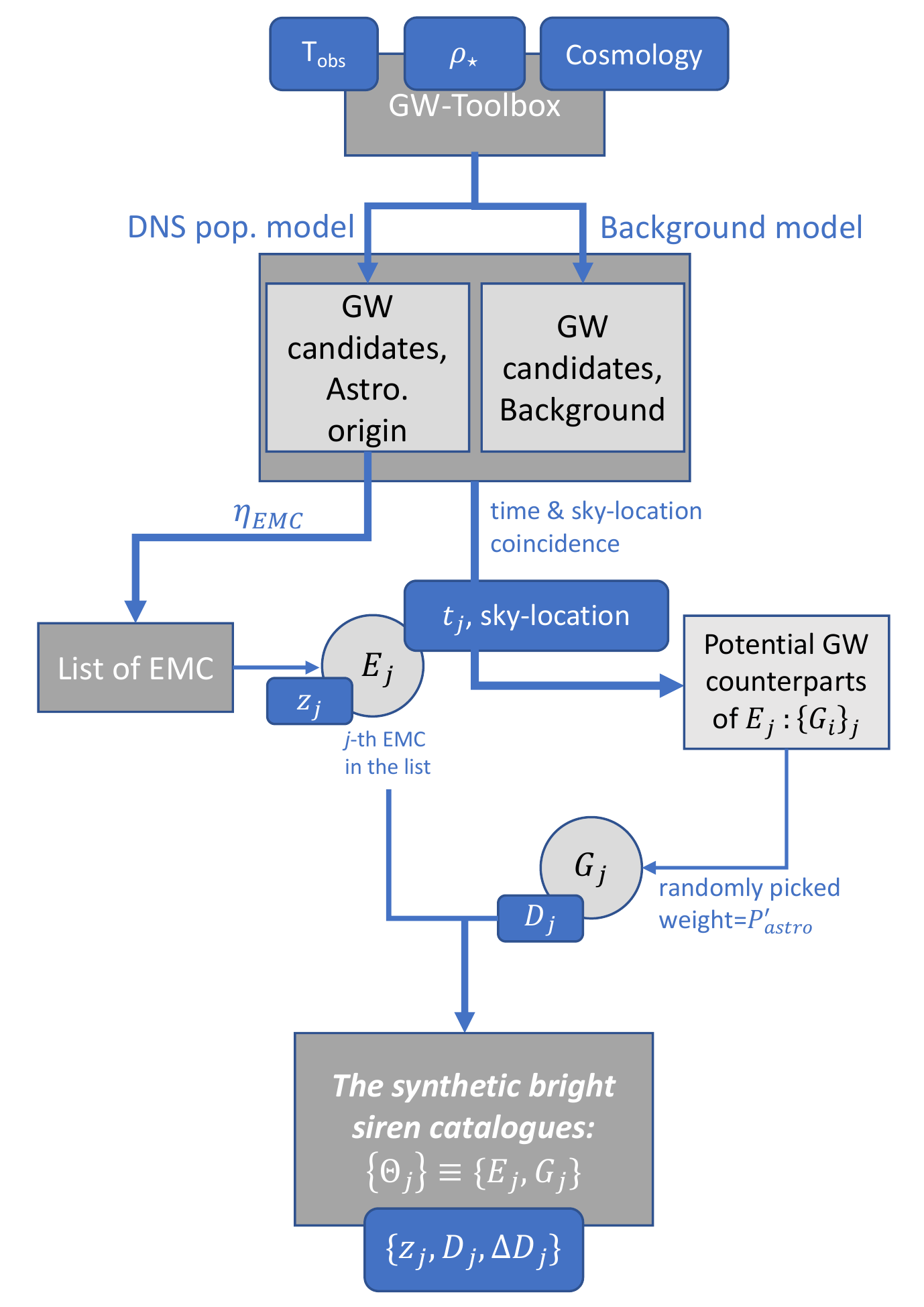}
    \caption{The sketch of Bright Siren. After the screening of the catalogues, we obtain the Bright Siren catalogues with red-shift from EMC and the $\rho$, $D$ and $\Delta D$ from GW sources.}
    \label{fig:sketch}
\end{figure}

So far, we have obtained a simulated catalogue of GW and their corresponding EMC, which we refer to as the bright siren catalogue. In Figure \ref{fig:sketch} we summarize our steps of simulating the bright siren catalogue. For our purpose of cosmology model constraints, the relevant parameters of these sources are their $D_{\rm L}$ and $z$. The former are attributes of the GW candidates, and the latter are of those EMC. Now $D_{\rm L}$ has a definitive value from simulation, whereas that from an observation can be shifted around this centre value according to its probability distribution. Here we assume the observed value of $D_{\rm L}$ follows a log-normal distribution \citep{2016PhRvD..93h3511O}, which is truncated at 1000\,Mpc (as the detecting limit of DNS). From here on, the values of $D_{\rm L}$ of the bright sirens are replaced by these re-sampled values according to this probability distribution.
%For simulating the real events, we assume the real distance according to the truncated log-normal distribution \citep[]{2016PhRvD..93h3511O}, which the mean is the original distance, the standard deviation is the original distance-error, the upper limit of truncation is 1000\,Mpc (which consider the furthest GW observed by LIGO). 
%There may be astronomical sources and background sources of the misidentification. So some data may significant deviations from model. We sifted out these data by limiting the relationship between distance, distance-error and red-shift to satisfy the Hubble constant greater than 60\,\rm{km\,s$^{-1}$\,Mpc$^{-1}$} and less than 80\,\rm{km\,s$^{-1}$\,Mpc$^{-1}$}. 

\begin{figure*}
    \centering
    \includegraphics[width=1.0\textwidth]{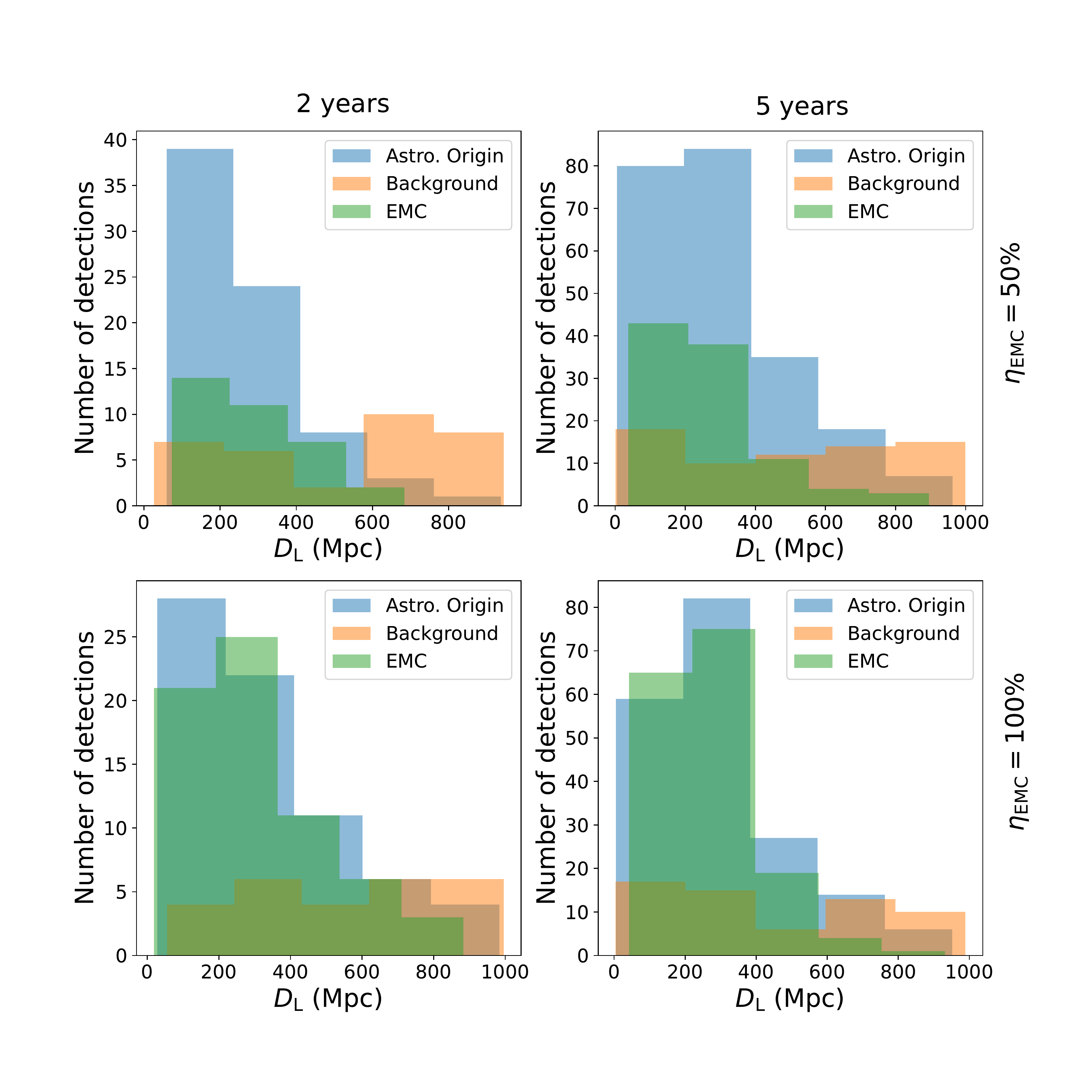}
    \caption{The histogram of luminosity distance for a set of simulated catalogues with $\rho_{\star}=6$, different $T_{\rm obs}$ and $\eta_{\rm EMC}$. The blue, orange and green represent astronomical originated GW sources, background GW sources and the EMC of the GW sources.}
    \label{fig:D_L}
\end{figure*}

In Table \ref{tab:table1}, we list some information of the simulated catalogues with various $T_{\rm obs}$, $\eta_{\rm EMC}$ and $\rho_\star$. The information includes the numbers of GW candidates of astronomical and background origins, the number of EMC and that of cases of mis-pairing, in one realization of simulation. In Figure \ref{fig:D_L}, we present the histogram of $D_{\rm L}$ of GW candidates of astronomical and background origins, and those of the GW candidates associated with EMC, of different catalogues. From Figure \ref{fig:D_L} one can see that the astronomical GW candidates clearly possess a different distribution of $D_{\rm L}$ than those of noises, indicating their different origins. While the distribution of EMC share similar shapes of those of the astronomical sources, implying that the mis-paring is rare, which is in agreement with Table \ref{tab:table1}. 

\begin{figure*}
    \centering
    \includegraphics[width=1.0\textwidth]{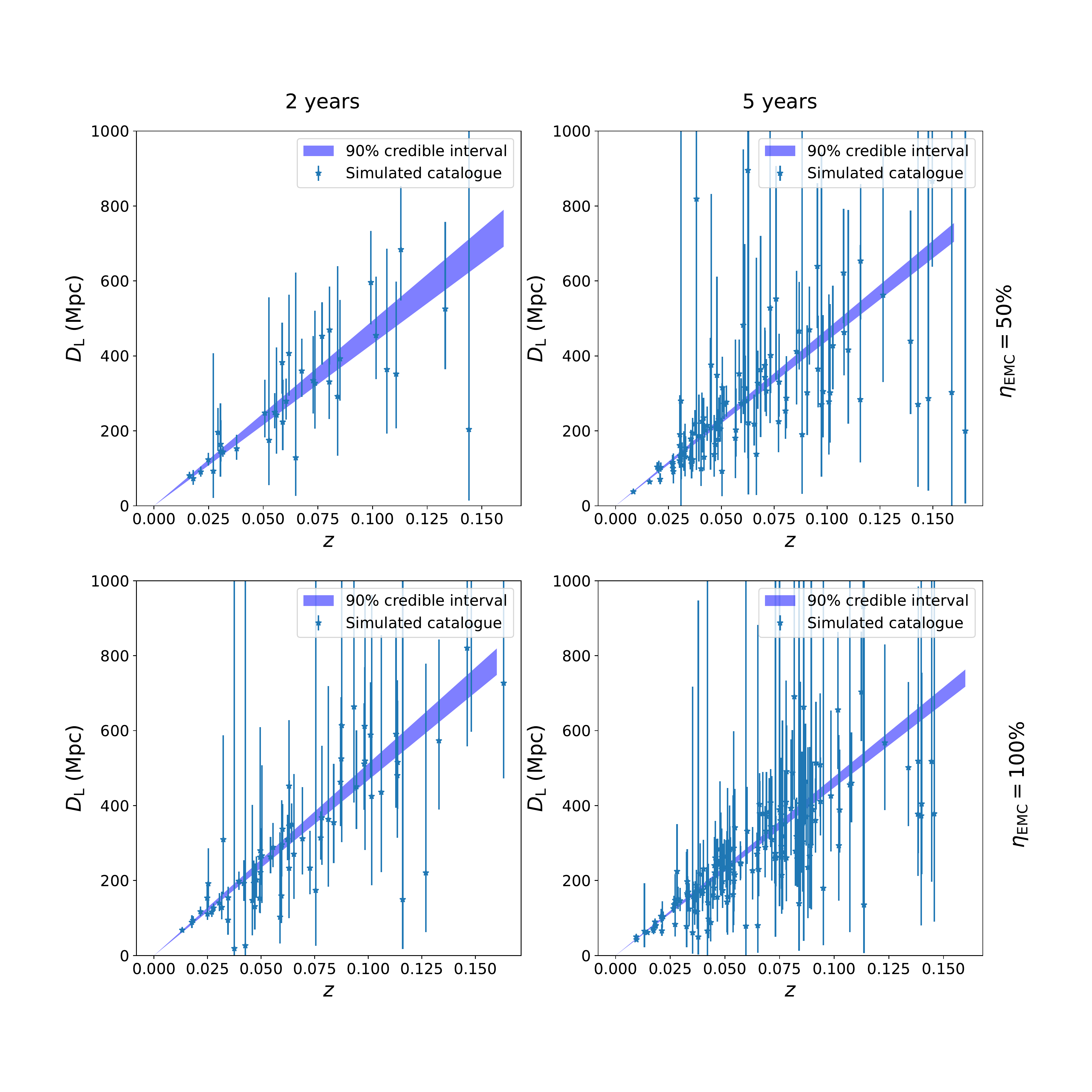}
    \caption{The $D_{\rm L}$ versus red-shift for a set of simulated catalogues with $\rho_{\star}=6$, various $T_{\rm obs}$ and $\eta_{\rm EMC}$. The constrain of $H_0$ is shown in the blue region in 90$\%$ credible interval. The points in the figure is the simulated catalogues used to constrain the $H_0$.}
    \label{fig:D_Z}
\end{figure*}

In Figure \ref{fig:D_Z}, we plot the $D_{\rm L}$ vs. $z$ diagrams of the catalogues from different settings of the simulation. The blue bands are the corresponding Bayesian inferred $D_{\rm L}-z$ relationship, which will be introduced in the following section.

%\sout{Based on figure \ref{fig:2figures}, we set 6 as the $\rho_\star$ to produce the initially GW events by \GWT\, and aLIGO-design background model. Then we get the sub-threshold bright-siren catalogues. For smaller $\rho_\star$ catalogues, we select them from these catalogues, which will improve the efficiency of the simulation. In this work, we set the $\rho_\star$ to 6, 7, 8, 9.

%The sketch of how to get the Bright Siren is shown in the figure \ref{fig:sketch}, where the $\rho_{\star}=6$, the $\eta_{\rm EMC}=50\%, 100\%$. Talbe \ref{tab:table1} shows the information of one Bright Siren catalogue, which include the conditions of different $\rho_\star$, observation duration and $\eta_{\rm EMC}$. The luminosity distance distribution and the constraint of $H_0$ (90$\%$ credible interval) in the luminosity distance vs red-shift figures, which the $\rho_\star$ is 6, are shown in figure \ref{fig:D_L} and figure \ref{fig:D_Z}, respectively.}

\begin{table*}
	\centering
	\begin{threeparttable}
	\caption{One representative set of different sub-threshold number of GW of astronomical origins, that of noises, the number of EMC, and the number of misidentification.}
	\label{tab:table1}
	\begin{tabular}{c|c|ccccc}
		\hline
		$T_{\rm obs}$ (years)  & $\eta_{\rm EMC}$ ($\%$) & $\rho_\star$ & $N_{\rm astro}$  & $N_{\rm background}$ & $N_{\rm EMC}$ & $N_{\rm misidentification}$ \\
		\hline
		\multirow{8}{*}{2} & \multirow{4}{*}{50} & 6 & 75 & 33 & 39 & 1   \\
		 &  & 7 & 48 & 0 & 23 &  0   \\
		 &  & 8 & 29 & 0 & 12 &  0  \\
		 &  & 9 & 22 & 0 & 8 &  0  \\ \cline {2-7}
		 & \multirow{4}{*}{100} & 6 & 71 & 26 & 71 & 0   \\ 
		 &  & 7 & 37 & 0 & 37 & 0   \\
		 &  & 8 & 28 & 0 & 28 & 0   \\
		 &  & 9 & 20 & 0 & 20 & 0   \\
		\hline \multirow{8}{*}{5} & \multirow{4}{*}{50} & 6 & 225 & 69 & 114 & 7  \\
		 &  & 7 & 158 & 0 & 81 & 2  \\
		 &  & 8 & 117 & 0 & 61 & 1  \\
		 &  & 9 & 90 & 0 & 47 & 0  \\ \cline {2-7}
		 & \multirow{4}{*}{100} & 6 & 188 & 61 & 188 & 2  \\
		 &  & 7 & 128 & 0 & 128 & 0  \\
		 &  & 8 & 96 & 0 & 96 & 0  \\
		 &  & 9 & 71 & 0 & 71 & 0  \\

		\hline
	\end{tabular}
   % \begin{tablenotes}
    %   \footnotesize
	%   \item - = unknown. 
    %\end{tablenotes}
    \end{threeparttable}
\end{table*}

\section{Bayesian inference on the $H_0$ from contaminated catalogues}

From those generated simulated bright siren catalogues with potential contamination with fake signals and mis-identification of EMC, we use a Bayesian method to estimate cosmological parameters. By Bayes theorem, the posterior distribution of cosmological parameters $\mathcal{B}$ is:
\begin{equation}
    p(\mathcal{B}|\{\Theta_i\})\propto\mathcal{L}(\{\Theta_i\};\mathcal{B})p(\mathcal{B}),\label{eq:Bayes}
\end{equation}
where $\{\Theta_i\}$ denotes the data in the catalogue, $\mathcal{L}(\{\Theta_i\};\mathcal{B})$ is the likelihood of obtaining data $\{\Theta_i\}$ supposing the parameters of the model is $\mathcal{B}$, and $p(\mathcal{B})$ is the prior probability distribution of $\mathcal{B}$. 

For the $i$-th entry in the simulated catalogue $\{D_{i,\rm{obs}},\Delta D_i,z_i\}$, the likelihood is:
\begin{equation}
    L_i=\int p(D_{i,\rm{true}}|D_{\rm{L},\rm{Theory}}(z_i;\mathcal{B}))p(D_{i,\rm{true}}|D_{i,\rm{obs}})dD_{i,\rm{true}}
\end{equation}
where $D_{\rm{L},Theory}$ the theoretical expected luminosity distance corresponding to red-shift $z_i$, given the cosmological model parameters $\mathcal{B}$. $p(D_{i,\rm{true}}|D_{\rm{L},\rm{Theory}}(z_i;\mathcal{B}))$ is the probability of getting $D_{\rm{i},true}$ given the expected value from theory is $D_{\rm{L},Theory}(z_i)$; $p(D_{i,\rm{true}}|D_{i,\rm{obs}})$ is the probability that the real luminosity distribution is $D_{i,\rm{true}}$, given the observed value in the catalogue is $D_{i,\rm{obs}}$. Both the above mentioned probability distribution are assumed to be normal in logarithm space \citep[]{2016PhRvD..93h3511O}, with the means at $\log D_{\rm{L},Theory}(z_i)$ and $\log D_{i,\rm{obs}}$ respectively, the standard deviation of $\Delta\log D_{i}=\Delta D_i/D_i$, and truncated at a maximum possible horizon at $D_{\rm{L},up}=1000$\,Mpc. 

The above integral can be approximated by the average over an ensemble of $D_{ij}$ where the index $j$ runs from 1 to $N_i$:

\begin{equation}
    L_i\approx\sum^{N_i}_{j=1}p(D_{ij,\rm{true}}|D_{\rm{L},\rm{Theory}})/N_i.
\end{equation}
The ensemble $D_{ij}$ is sampled from a $p(D_{ij}|D_{i,\rm{obs}})$. In our study, we take $N_i=50$. 

The likelihood for obtaining a catalogue $\{\Theta_i\}$, where $\Theta_i=\{D_{i,\rm{obs}},\Delta D_i,z_i\}$, is:
\begin{equation}
    \mathcal{L}(\{\Theta_i\};\mathcal{B})=\prod^N_{i=1}L_i(\Theta_i;\mathcal{B}).
    \label{eq:likel4}
\end{equation}

In this work, we take $D_{\rm{L},Theory}(z;\mathcal{B})$ as modeled by the flat $\Lambda$-CDM cosmological model, whose parameters are $H_0$, $\Omega_\Lambda$ and $\Omega_m$, and $\Omega_\Lambda+\Omega_m=1$. At low redshift approximation, 
\begin{equation}
    D_{\rm{L},Theory}(z)\approx\frac{cz}{H_0},
    \label{eq:Dtheory}
\end{equation}
where $c$ is the speed of light. 

We use a Gaussian distribution prior of $H_0$, which centers at 70\,\rm{km\,s$^{-1}$\,Mpc$^{-1}$} with a standard deviation of 40\,\rm{km\,s$^{-1}$\,Mpc$^{-1}$}. When considering the higher order contribution with a flat $\Lambda$-CDM cosmology, the prior of $\Omega_m$ is assumed to be uniform between 0 and 1. We use a Markov Chain-Monte Carlo (MCMC) algorithm to obtain a sample from the resulted posterior distribution with Equation \ref{eq:Bayes}. 

\begin{figure}
    \centering
    \includegraphics[width=.5\textwidth]{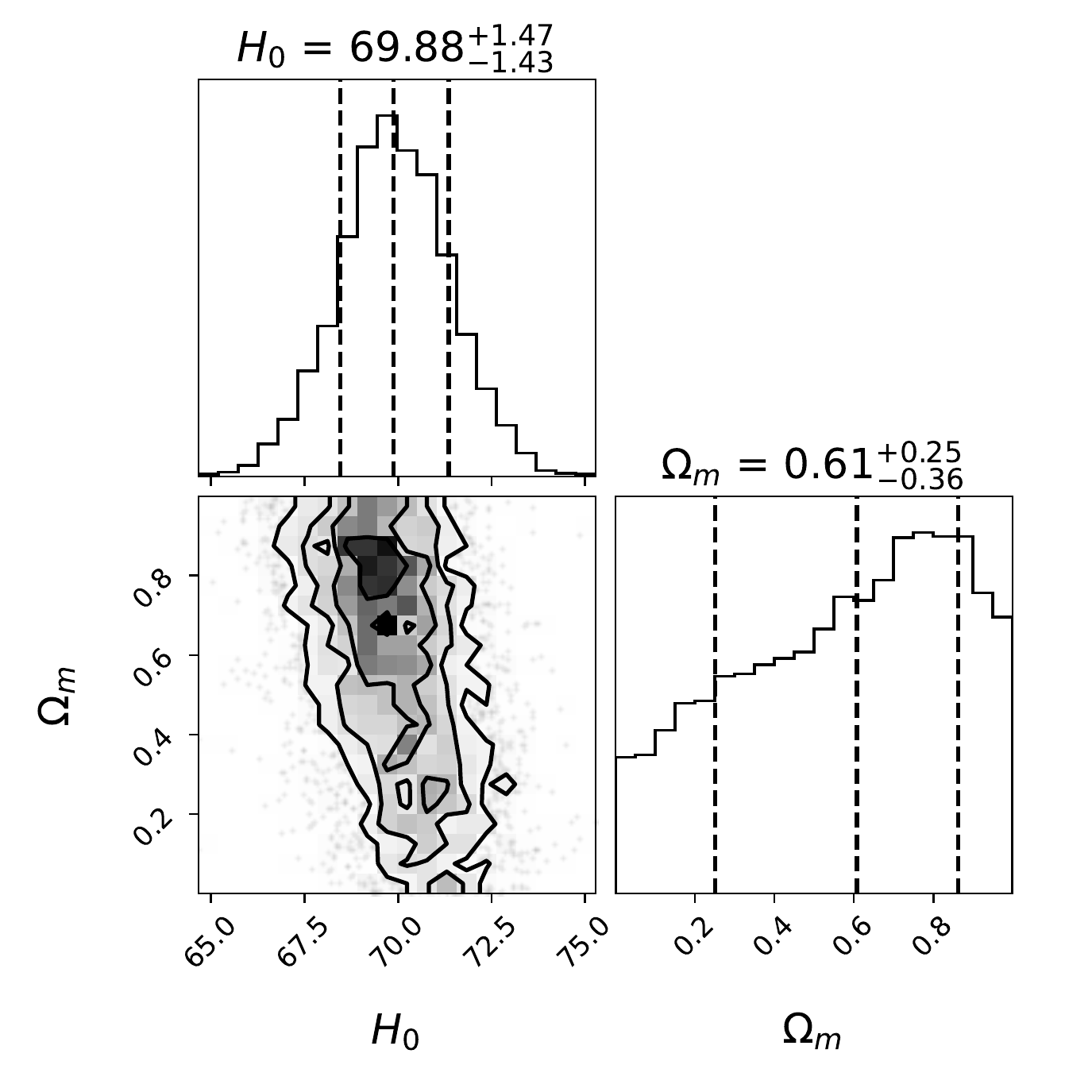}
    \caption{The constraints on $H_0$ and $\Omega_{m}$ in Flat$\Lambda$CDM model by using a $\rho_\star$=6, $T_{\rm obs}=5$ years and $\eta_{\rm EMC}=100 \%$ catalogue, which ensures that there are enough bright sirens in the catalogue. The three vertical black dashed lines is the 16$\%$, 50$\%$ and 84$\%$ percentiles.}
    \label{fig:Flat}
\end{figure}

As an example, in Eigure \ref{fig:Flat} we plot the probability distributions of the priors of $H_0$ and $\Omega_m$ respectively, and the contours (showing the 68\% quantiles) of their covariance, which are inferred with an synthetic catalogue of 5 years observation, 100$\%$ $\eta_{\rm EMC}$ and $\rho_\star=6$. This catalogue corresponds to the lower-right panel of Figure \ref{fig:D_Z}.  As we can find in Figure \ref{fig:Flat}, although the sub-threshold catalogue includes events as far as $z\sim0.15$, the higher order effect in the $z-D_L$ relation is still not significant. As a result of this, $\Omega_m$ cannot be constrained meaningfully. Therefore, in the rest of our investigation, we adopt the low red-shit approximation of $D_{\rm{L,Theory}}(z)$ as in Equation \ref{eq:likel4}. Thus, the only cosmological parameter left to infer is $H_0$. 
 
%We use the 5 years and 6 S/N threshold to ensure the bright siren catalogue large enough. figure \ref{fig:Flat} shows that we can not constrain the $\Omega_m$ in flat $\Lambda$-CDM model by sub-threshold GW catalogues.

\section{Results and conclusion}

\begin{figure*}
    \centering
    \includegraphics[width=1.0\textwidth]{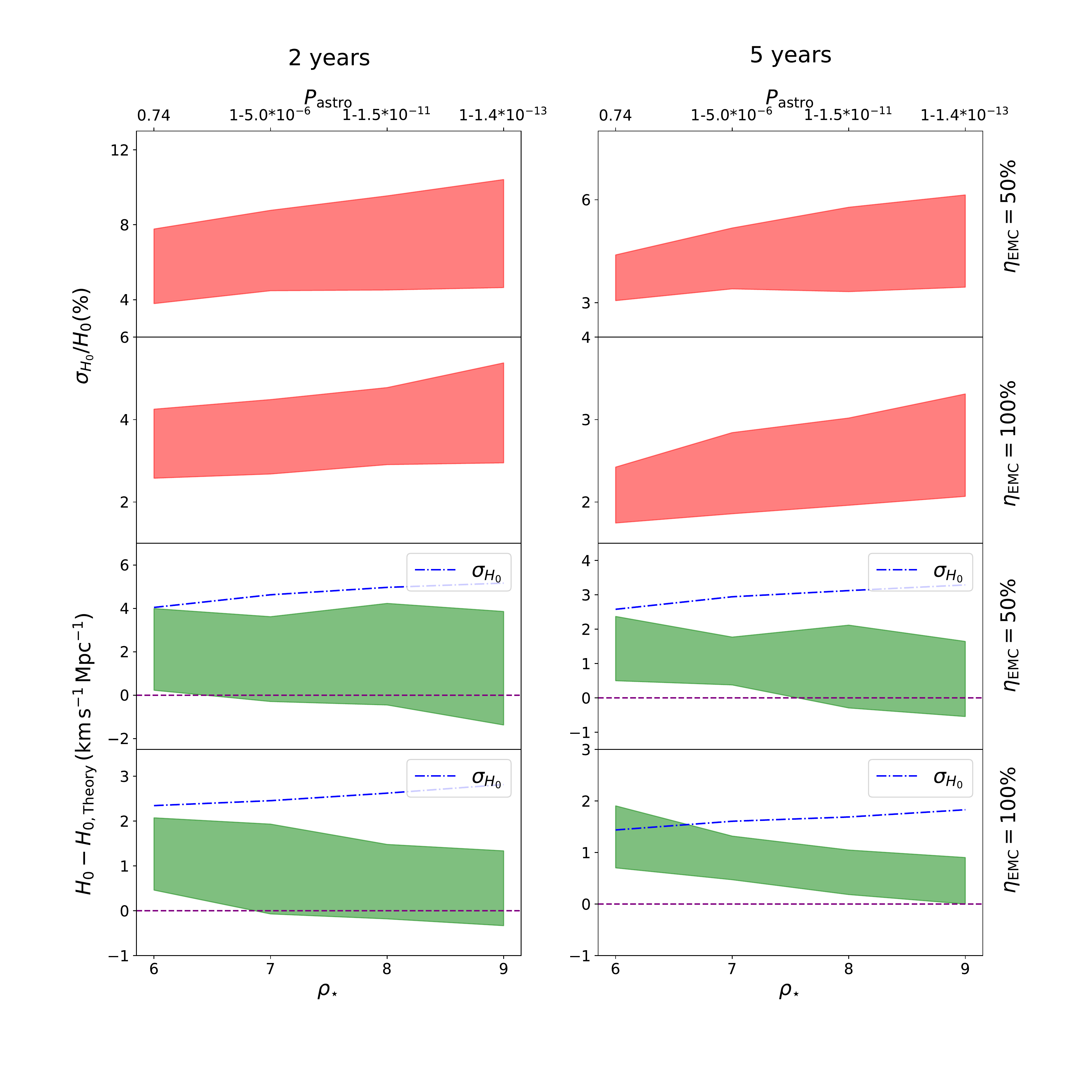}
    \caption{The $H_0$ fractional measurement error (two top panels), defined as $\sigma_{H_0}$ divided by the posterior median of $H_0$, and $H_0$ measurement bias (two bottom panels) as a function of the $\rho_\star$ (90$\%$ credible interval). The first and third panels are the results of $\eta_{\rm EMC}$ is 50$\%$, the second and forth panel are the results of $\eta_{\rm EMC}$ is 100$\%$.  The 2 years and 5 years results are shown in left and right, respectively. The upper axis is $P_{\rm astro}$ corresponding to $\rho_\star$. Here, $H_{0,\rm Theory}$ is 67.4\,\rm{km\,s$^{-1}$\,Mpc$^{-1}$} \citep[]{2020A&A...641A...6P}. We also plot the $\sigma_{H_0}$ (blue dash dotted line), which is the average of the upper and lower bounds in two top panels, in the two bottom panels.}
    \label{fig:results}
\end{figure*}

A realization of simulated catalogue will get a posterior of $H_0$. The blue region in Figure \ref{fig:D_Z} represents the posterior of $H_0$ (90$\%$ credible interval). We get $D_{\rm L}$ by $D_{\rm L, i}=c\times z_i/H_{0,i}$, where $c$ is the velocity of light, $z_i$ is the value selected with the equal step from 0 to 0.16, $H_{0,i}$ is the 5$\%$ percentiles and 95$\%$ percentiles of the posterior of $H_0$. Using the values of $D_{\rm L, i}$ and $z_i$, we plot the blue region as the 90$\%$ credible interval in Figure \ref{fig:D_Z}.
The $H_0$ measurement error ($\sigma_{H_0}$) is defined as half of the width of the 68$\%$ symmetric credible interval of the posterior of $H_0$ divided by the posterior median. The $H_0$ measurement bias is defined as the posterior median minus the theoretical value, which is the Plank18's measurement result.   
We make many realizations of simulated catalogues with the same setting of the $T_{\rm obs}$, $\rho_{\star}$ and $\eta_{\rm EMC}$. For each realization, we calculate the $H_0$ measurement error and the $H_0$ measurement bias. After that, we obtain many $\sigma_{H_0}$ and the bias of $H_0$, from which we can find a 90$\%$, credible interval as the upper and lower limits as in Figure \ref{fig:D_Z}. We plott these region as function of $\rho_{\star}$ for different $T_{\rm obs}$ and $\eta_{\rm EMC}$ in Figure \ref{fig:results}. The number of realizations we apply in this study is 30, which we find is enough to give robust upper and lower limits of the $\sigma_{H_0}$ and the bias of $H_0$.   

%Fig \ref{fig:results} shows the $H_0$ measurement error and $H_0$ measurement bias with different observation times and $\eta_{\rm EMC}$. Fig \ref{fig:results} shows that the error increases with the increase of $\rho_\star$ or $P_{\rm astro}$ and decreases with the increase of the $\eta_{\rm EMC}$. The lower limit of the error rises slowly as the $\rho_\star$ or $P_{\rm astro}$ increase. The minimum error is about 3.5$\%$ or 2$\%$ for 2 or 5 years. The bias increase with the $\rho_\star$ or $P_{\rm astro}$ increase. In most cases, the bias is less than 1 $\sigma_{H_0}$ and is positive, which means that the posterior $H_0$ is greater than the theory value.      

%We found that when the gravitational wave detection threshold is lowered to a critical level ($P_{\rm astro}$ $\sim$ 0.7), despite the unfavorable factors such as misidentification of EMC, false GW signals, and greater parameter uncertainty in multi-messenger catalogues, the constraints on the $H_0$ is better by more bright sirens.

%The lower limit of $H_0$ measurement error varies less with the $\rho_\star$. Therfore, the constraints on H0 is mainly determined by part data with a very high $\rho$.

%When we use the FlatLambdaCDM model, we find that the matter density parameter $\Omega_m$ can't be constrained by the sub-threshold catalogues. Although, the red-shift of the GW events reach $\sim$ 0.15.
We conclude as follows:
\begin{itemize}
    \item As long as $P_{\rm{astro}}\gtrsim0.1$, lower the $\rho_\star$ will significantly enlarge the bright-siren candidates catalogue, while the fraction of contamination due to mis-identification and background noise is small (see table \ref{tab:table1}).
    \item The marginal candidates in the sub-threshold catalogue are those with higher red-shift and uncertainties. Inclusion of those populations will on average tighten the constraint on $H_0$ (upper panels of figure \ref{fig:results}), while leaves the higher order parameters such as $\Omega_m$ unconstrained. 
    \item The constraint of $H_0$ improves with lower $\rho_\star$, which can be seen as a general tendency in many realizations of the simulations. However, it is
    interesting to observe from Figure \ref{fig:results} that such tendency is more prominent among those realizations which result in poorest $H_0$ constraints.
    \item In the lower panels of Figure \ref{fig:results}, we can also find that, with lower $\rho_\star$, the inferred $H_0$ tends to bias towards greater values. We discuss on its origin in the next section.
\end{itemize}
\section{discussion }
%\sout{The things we can improve.}
\subsection{A network of GW detectors}

There is an obvious distinction between our simulation study and a realistic case: our simulation is for a single GW detector, while in reality GW observations are in general made by a network of detectors. This distinction has some fundamental influence on our results quantitatively. One major influence is on the sky-localization estimation, which plays a role in this study in a double-folded way: first, the mis-pairing probability is determined by the localization of candidates; second, in a standard Bayesian parameter estimation procedure, the uncertainties of the luminosity distance are in degeneration with the sky coordinates, and the former will directly enter the constraints on $H_0$. The localization ability of a network of GW detectors is fundamentally different from a single detector. Since the current \GWT\, is not able to simulate an observation with a network, and the sky coordinates uncertainties given by \GWT\, with a FIM based on a single detector will be largely overestimated compared to a more realistic detector network, we instead use the localization estimation from the interpolation of an empirical relation between $\Delta\Omega$ and $\rho$, which is fitted from previous GW catalogues.

The ability to simulate the observation with a network is being added into the \GWT\, as a major update. The localization of the network will be estimated with two implementations: the first one based on triangulation where only the times of arrival of the signals at all detectors are used \citep{2011CQGra..28j5021F,2020LRR....23....3A}; the second one also makes use of the amplitudes and phases of the signals upon arrival \citep{2017PhRvD..95d2001M,2018CQGra..35j5002F}. 

The substitution of a single detector simulation with a network will have the following quantitative influence on our results: first, the sky coordinates uncertainties will be estimated more accurately for individual candidates, and will be in general smaller for future network than those obtained from interpolation with previous observations. The resulted mis-pairing rate will be therefore less than what is found in this simulation; the uncertainty estimation of the luminosity distance will be more accurate, where the overestimation of $\Delta D$ towards low $\rho$ expected in this study will be less serious. As a result, the $H_0$ constraints can be tighter. With the criteria of time-coincidence of signals, the background noise level is expected to be lower than that from a single detector, therefore $\rho_{\star}$ can be even lower when the target $P_{\rm astro}$ is set to the same.

%We only simulate the aLIGO-design background and the GW observation on DNS by \GWT. For a more realistic simulation, we should add the a network of GW detectors. The addition of network will not only improve the S/N, so increase the number of GW that can be detected, but also improve the space-time localization of GW signals. We can decrease the luminosity distance error $\Delta D$ and the uncertainties of sky-localization $\Delta\Omega$ though adding the network, thereby improving the accuracy of the simulation. We are trying adding the network in the \GWT, two method is considered: one only consider the timing-triangulation alone \citep{2011CQGra..28j5021F,2020LRR....23....3A}, while the other one takes into consider the amplitudes and phases of the signal \citep{2017PhRvD..95d2001M,2018CQGra..35j5002F}.    

\subsection{The EMC follow-up efficiency}
As introduced in the above sections, well-known EMCs of DNS mergers include GRB (prompt emission and afterglow), optical and radio afterglow and kilonovae etc.. In this paper, we do not specify the EMC of GW, that is, any EMC follow-up observation campaign which leads to GW's independent $z$ measurement will fit with our study. However, the EMC follow-up efficiency $\eta_{\rm EMC}$, which plays an important role in our study, does depend on the class of EMC observation targets. In this paper, we set the fiducial $\eta_{\rm EMC}$ with 50$\%-100\%$, which roughly agrees with those of kilonovae. For a careful estimation of $\eta_{\rm EMC}$ of a kilonova, we should take account of the variation of the efficiency with the source distance, kilonova models, filters and so on. From \cite{2021MNRAS.504.1294S}'s simulation of the kilonova survey, we found that kilonova models, geometry, viewing-angle, wavelength coverage, source distance and the minimum last post-merger epoch required will affect the detection probability. The kilonova models' some parameters can be constrained by the AT 2017gfo, such as the total ejecta mass and the half-opening angle ($\phi$, a lanthanide-rich component distributed within $\pm \phi$ around the merger/equatorial plane \citep[]{2019MNRAS.489.5037B}). The limiting magnitude and distance is different for different optical telescopes. We should give a uniform maximum limiting magnitude and maximum kilonova detection distance. To simplify the problem, we can unify the wavelength coverage, i.e., the filters, and give the distribution to the viewing-angle and the minimum last post-merger epoch required. We can then simulate the kilonova follow-up efficiency, and improve our simulation of the EMC of GW.

In reality, EMC follow-up will largely depend on the  multi-wavelength observation. Early information from high energy bands will largely contribute to the improvement of $\eta_{\rm EMC}$. In the example of GRBs, their prompt emission will be easier to be recognized, if they are simultaneous in time with GW (time lag $\sim$ s). In the meanwhile, they can provide the positioning information of the GW, which increases the multi-wavelength follow-up efficiency. However, the coincidence fraction between a GW and GRB prompt emission is expected to be low, due to the highly collimated emission beam of GRB prompt emission \citep{2022arXiv220814156H}. However, if we consider the precursor and afterglow of a GRB, the coincidence fraction will increase. Similarly, they will also improve the efficiency of multi-wavelength follow-up. In addition, there is a recent model proposed in which a DNS merger involving a magnetar will have gamma-ray precursor \citep{2022ApJ...939L..25Z}, which also is helpful to improve multi-wavelength follow-up efficiency. %we propose that a part of DNS mergers may have gamma-ray precursor\citep{2022ApJ...939L..25Z}, which also is helpful to improve EMC follow-up efficiency.   

\subsection{Bias of the $H_0$ posterior}
\begin{figure*}
    \centering
    \includegraphics[width=1.0\textwidth]{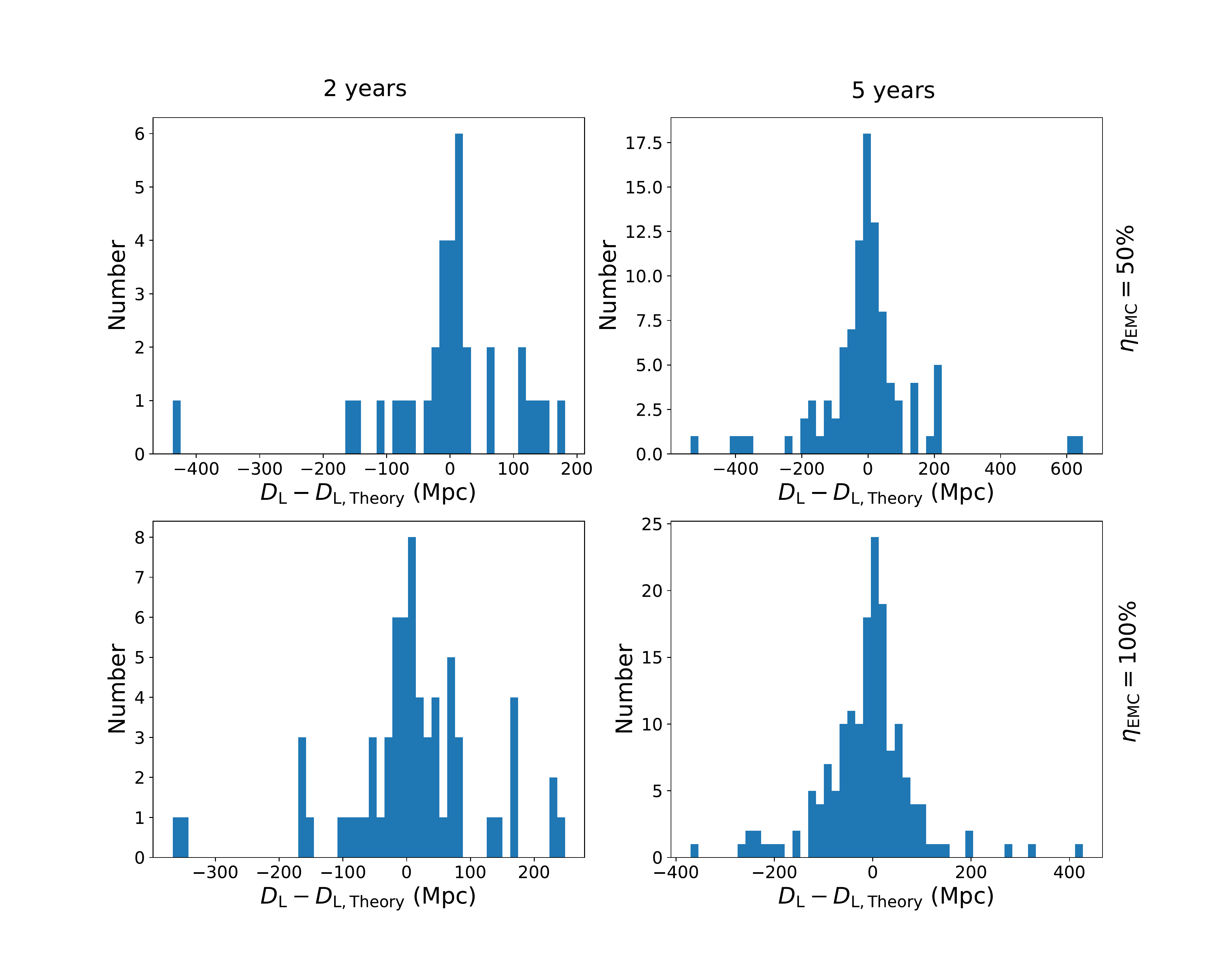}
    \caption{The distribution of $D_{\rm L}-D_{\rm L, Theory}$ corresponds to figure \ref{fig:D_Z}. The $D_{\rm L, Theory}$ is calculated by equation \ref{eq:Dtheory}, where $H_0$ is the $H_{0,\rm Theory}$.}
    \label{fig:DLDT}
\end{figure*}

As can be seen from the lower panels of Figure \ref{fig:results}, the inferred $H_0$ tends to have a higher value compared with the one which was used to generate the data. The discrepancies between the inferred ones and the injected $H_0$ are in most cases less than $\sigma_{H_0}$, which means the biases are not as significant as the 90\% confident level. The biases are generally larger when the catalogues have smaller $\rho_{\star}$, or equivalently less $P_{\rm {astro}}$. This tendency is partly due to the decrease of $\sigma_{H_0}$ inclination towards smaller $\rho_{\star}$, while the absolute difference between the inferred and injected also increases. The latter is attributed to the bias of $D_{\rm L}$ for the marginal candidates in the catalogue. As mentioned in the above section, in the calculation of the likelihood in the Bayesian inference, $D_{\rm L}$ is re-sampled over a probability distribution of the true $D_{\rm L}$. The probability distribution of $D_{\rm L}$ is assumed to be a log-normal distribution, sharply truncated at 1000\,Mpc, which corresponds to the detection limit. Those marginal candidates in a sub-threshold catalogues have their $D_{\rm L}$ close to the detection limit, and their $\Delta D$ inferred with FIM are also likely to be as large as in the same order of magnitude of $D_{\rm L}$. For those candidates, a re-sampling over a truncated log-normal distribution will result $D_{\rm L}$ asymmetric  and likely to have smaller values than their genuine ones. In Figure \ref{fig:DLDT}, we plot the histograms of the difference between the re-sampled $D_{\rm L}$ and the corresponding genuine one, which is $cz/H_{0,\rm Theory}$. We can see the asymmetry of the difference towards smaller $D_{\rm L}$. Such biases of $D_{\rm L}$ of marginal candidates towards smaller values results in a $H_0$ posterior biases towards large values. 

This finding from the simulation study may reflect a basic issue of the sub-threshold standard siren methodology: the probability distribution of $D_{\rm L}$ will be highly asymmetrical for those marginal candidates, and thus results in biased estimation of $H_0$, which will systematically towards larger values. However, we expect that with an accurate estimation of the $D_{\rm L}$ true posterior distribution from the Bayesian method, rather than the rough methods applied here, such biases can be reduced. On the other hand, with a more realistic stimulation study, such potential bias in $H_0$ can be known and corrected. Such simulation studies will be crucial if the standard siren method is used to arbitrate the Hubble tension.
Besides the aforementioned effect which causes the bias of $H_0$ in sub-threshold catalogues, there are two other sources of biases, which will inflate the inferred value $H_0$ as discussed in literature. One originates from a combination effect of the inclination-distance degeneracy and the selection favorite over small inclination angle \citep{2021PhRvD.104h3531G}; the other effect is due to the selection bias of the EMC towards small inclination angle \citep{2020PhRvL.125t1301C}. Both effects will result in a skewed luminosity distance posterior, and thus bias the $H_0$. \cite{2023MNRAS.520....1G} provided a method which can correct the biases from the above-mentioned two effects. In our study, these two effects do not contribute to the $H_0$ bias, since we use a presumed $D_{\rm L}$ posterior. As discussed above, the source of $H_0$ bias is a separated one, arising from the events near the detection horizon. It therefore becomes dominating in the case of using sub-threshold catalogue, where the number of events near the horizon is large.
%张双南评论：我觉得最后一句话有问题。由于bias系统的增加H0，marginal candidates的数量越多，bias越严重，并不会由于数量增加而把bias平均掉。
%The $H_0$ posterior is greater than the theory value in most cases, the amplitude is basically less than $1\sigma_{H_0}$. The reason for that is the small $D_\rm{L}$, as shown in the figure 9, $D_{\rm{L}}-\frac{cz}{H_{0,\rm{Theory}}}$ are more distributed on the side less than zero. We assume that the $D_{\rm{L}}$ conforms to the truncated log-normal distribution, with an upper bound of 1000\,Mpc, which is the detection distance limit of GW detectors. When generating sub-threshold catalogues,signals with a low S/N have a large $\Delta D$. The $D_{\rm{L}}$ of these GW signals are likely to be greater than 1000\,Mpc, however, the truncated log-normal distribution will shrink $D_{\rm{L}}$ range to less than 1000\,Mpc. That will cause the $D_{\rm{L}}$ to be smaller than theoretically predicted. At the same time, the lower the $\rho_\star$, the larger the bias of the $H_0$ posterior. 

%The bias of the $H_0$ posterior in the simulation will reflect the real problem in the detective. The detection distance limit of the detector will cause that the distribution of the $D_{\rm{L}}$ near the limit is extremely asymmetrical, with much more values below the limit than values greater than the limit. Therefore, the $D_{\rm{L}}$ will smaller than the theoretically predicted in the GW observation. The problem is more serious for the sub-threshold observation.    
\section*{Acknowledgments}
This work is supported by the National Key R\&D Program of China (2021YFA0718500). SXY acknowledges the support by the Institute of High Energy Physics (Grant No. E25155U1).


\begin{thebibliography}{}
\bibliographystyle{plain}
\bibitem[Abbott et al.(2017)]{2017Natur.551...85A} Abbott, B.~P., Abbott, R., Abbott, T.~D., et al.\ 2017, \nat, 551, 85. doi:10.1038/nature24471
\bibitem[Abbott et al.(2017)]{2017ApJ...848L..12A} Abbott, B.~P., Abbott, R., Abbott, T.~D., et al.\ 2017, \apjl, 848, L12. doi:10.3847/2041-8213/aa91c9
\bibitem[Abbott et al.(2020)]{2020ApJ...892L...3A} Abbott, B.~P., Abbott, R., Abbott, T.~D., et al.\ 2020, \apjl, 892, L3. doi:10.3847/2041-8213/ab75f5

%\bibitem[Abbott et al.(2020)]{2020CQGra..37e5002A} Abbott, B.~P., Abbott, R., Abbott, T.~D., et al.\ 2020, Classical and Quantum Gravity, 37, 055002. doi:10.1088/1361-6382/ab685e
\bibitem[Abbott et al.(2020)]{2020LRR....23....3A} Abbott, B.~P., Abbott, R., Abbott, T.~D., et al.\ 2020, Living Reviews in Relativity, 23, 3. doi:10.1007/s41114-020-00026-9
\bibitem[Bulla(2019)]{2019MNRAS.489.5037B} Bulla, M.\ 2019, \mnras, 489, 5037. doi:10.1093/mnras/stz2495
\bibitem[Chase et al.(2022)]{2022ApJ...927..163C} Chase, E.~A., O'Connor, B., Fryer, C.~L., et al.\ 2022, \apj, 927, 163. doi:10.3847/1538-4357/ac3d25
\bibitem[Chen et al.(2018)]{2018Natur.562..545C} Chen, H.-Y., Fishbach, M., \& Holz, D.~E.\ 2018, \nat, 562, 545. doi:10.1038/s41586-018-0606-0
\bibitem[Chen(2020)]{2020PhRvL.125t1301C} Chen, H.-Y.\ 2020, \prl, 125, 201301. doi:10.1103/PhysRevLett.125.201301

\bibitem[Del Pozzo(2012)]{2012PhRvD..86d3011D} Del Pozzo, W.\ 2012, \prd, 86, 043011. doi:10.1103/PhysRevD.86.043011
\bibitem[Fairhurst(2011)]{2011CQGra..28j5021F} Fairhurst, S.\ 2011, Classical and Quantum Gravity, 28, 105021. doi:10.1088/0264-9381/28/10/105021
\bibitem[Fairhurst(2018)]{2018CQGra..35j5002F} Fairhurst, S.\ 2018, Classical and Quantum Gravity, 35, 105002. doi:10.1088/1361-6382/aab675

\bibitem[Feeney et al.(2021)]{2021PhRvL.126q1102F} Feeney, S.~M., Peiris, H.~V., Nissanke, S.~M., et al.\ 2021, \prl, 126, 171102. doi:10.1103/PhysRevLett.126.171102
\bibitem[Fishbach et al.(2019)]{2019ApJ...871L..13F} Fishbach, M., Gray, R., Maga{\~n}a Hernandez, I., et al.\ 2019, \apjl, 871, L13. doi:10.3847/2041-8213/aaf96e
\bibitem[Gagnon-Hartman et al.(2023)]{2023MNRAS.520....1G} Gagnon-Hartman, S., Ruan, J., \& Haggard, D.\ 2023, \mnras, 520, 1. doi:10.1093/mnras/stad069
\bibitem[Gerardi et al.(2021)]{2021PhRvD.104h3531G} Gerardi, F., Feeney, S.~M., \& Alsing, J.\ 2021, \prd, 104, 083531. doi:10.1103/PhysRevD.104.083531
\bibitem[Ghosh et al.(2022)]{2022arXiv220311756G} Ghosh, T., Biswas, B., \& Bose, S.\ 2022, arXiv:2203.11756
\bibitem[Hendriks et al.(2022)]{2022arXiv220814156H} Hendriks, K., Yi, S.-X., \& Nelemans, G.\ 2022, arXiv:2208.14156
\bibitem[Hinshaw et al.(2013)]{2013ApJS..208...19H} Hinshaw, G., Larson, D., Komatsu, E., et al.\ 2013, \apjs, 208, 19. doi:10.1088/0067-0049/208/2/19
\bibitem[Holz \& Hughes(2005)]{2005ApJ...629...15H} Holz, D.~E. \& Hughes, S.~A.\ 2005, \apj, 629, 15. doi:10.1086/431341
\bibitem[Jang \& Lee(2017)]{2017arXiv170201118J} Jang, I.~S. \& Lee, M.~G.\ 2017, arXiv:1702.01118
\bibitem[Lynch et al.(2018)]{2018ApJ...861L..24L} Lynch, R., Coughlin, M., Vitale, S., et al.\ 2018, \apjl, 861, L24. doi:10.3847/2041-8213/aacf9f
\bibitem[MacLeod \& Hogan(2008)]{2008PhRvD..77d3512M} MacLeod, C.~L. \& Hogan, C.~J.\ 2008, \prd, 77, 043512. doi:10.1103/PhysRevD.77.043512
\bibitem[Messick et al.(2017)]{2017PhRvD..95d2001M} Messick, C., Blackburn, K., Brady, P., et al.\ 2017, \prd, 95, 042001. doi:10.1103/PhysRevD.95.042001

\bibitem[Metzger et al.(2010)]{2010MNRAS.406.2650M} Metzger, B.~D., Mart{\'\i}nez-Pinedo, G., Darbha, S., et al.\ 2010, \mnras, 406, 2650. doi:10.1111/j.1365-2966.2010.16864.x
\bibitem[Mortlock et al.(2019)]{2019PhRvD.100j3523M} Mortlock, D.~J., Feeney, S.~M., Peiris, H.~V., et al.\ 2019, \prd, 100, 103523. doi:10.1103/PhysRevD.100.103523
\bibitem[Nissanke et al.(2010)]{2010ApJ...725..496N} Nissanke, S., Holz, D.~E., Hughes, S.~A., et al.\ 2010, \apj, 725, 496. doi:10.1088/0004-637X/725/1/496
\bibitem[Oguri(2016)]{2016PhRvD..93h3511O} Oguri, M.\ 2016, \prd, 93, 083511. doi:10.1103/PhysRevD.93.083511
\bibitem[Phillips(1993)]{1993ApJ...413L.105P} Phillips, M.~M.\ 1993, \apjl, 413, L105. doi:10.1086/186970
\bibitem[Planck Collaboration et al.(2014)]{2014A&A...571A..16P} Planck Collaboration, Ade, P.~A.~R., Aghanim, N., et al.\ 2014, \aap, 571, A16. doi:10.1051/0004-6361/201321591
\bibitem[Planck Collaboration et al.(2016)]{2016A&A...594A..13P} Planck Collaboration, Ade, P.~A.~R., Aghanim, N., et al.\ 2016, \aap, 594, A13. doi:10.1051/0004-6361/201525830
\bibitem[Planck Collaboration et al.(2020)]{2020A&A...641A...6P} Planck Collaboration, Aghanim, N., Akrami, Y., et al.\ 2020, \aap, 641, A6. doi:10.1051/0004-6361/201833910
\bibitem[Rastinejad et al.(2022)]{2022ApJ...927...50R} Rastinejad, J.~C., Paterson, K., Fong, W., et al.\ 2022, \apj, 927, 50. doi:10.3847/1538-4357/ac4d34
\bibitem[Riess et al.(1995)]{1995ApJ...438L..17R} Riess, A.~G., Press, W.~H., \& Kirshner, R.~P.\ 1995, \apjl, 438, L17. doi:10.1086/187704
\bibitem[Riess et al.(2016)]{2016ApJ...826...56R} Riess, A.~G., Macri, L.~M., Hoffmann, S.~L., et al.\ 2016, \apj, 826, 56. doi:10.3847/0004-637X/826/1/56
\bibitem[Riess et al.(2022)]{2022ApJ...934L...7R} Riess, A.~G., Yuan, W., Macri, L.~M., et al.\ 2022, \apjl, 934, L7. doi:10.3847/2041-8213/ac5c5b
\bibitem[Sagu{\'e}s Carracedo et al.(2021)]{2021MNRAS.504.1294S} Sagu{\'e}s Carracedo, A., Bulla, M., Feindt, U., et al.\ 2021, \mnras, 504, 1294. doi:10.1093/mnras/stab872
\bibitem[Schutz(1986)]{1986Natur.323..310S} Schutz, B.~F.\ 1986, \nat, 323, 310. doi:10.1038/323310a0
\bibitem[Spergel et al.(2007)]{2007ApJS..170..377S} Spergel, D.~N., Bean, R., Dor{\'e}, O., et al.\ 2007, \apjs, 170, 377. doi:10.1086/513700
\bibitem[The LIGO Scientific Collaboration et al.(2021)]{2021arXiv211103606T} The LIGO Scientific Collaboration, the Virgo Collaboration, the KAGRA Collaboration, et al.\ 2021, arXiv:2111.03606
\bibitem[Verde et al.(2019)]{2019NatAs...3..891V} Verde, L., Treu, T., \& Riess, A.~G.\ 2019, Nature Astronomy, 3, 891. doi:10.1038/s41550-019-0902-0
\bibitem[Wang et al.(2003)]{2003ApJ...590..944W} Wang, L., Goldhaber, G., Aldering, G., et al.\ 2003, \apj, 590, 944. doi:10.1086/375020
\bibitem[Xiao et al.(2022)]{2022arXiv220502186X} Xiao, S., Zhang, Y.-Q., Zhu, Z.-P., et al.\ 2022, arXiv:2205.02186
\bibitem[Yi et al.(2021)]{2021arXiv210613662Y} Yi, S.-X., Nelemans, G., Brinkerink, C., et al.\ 2021, arXiv:2106.13662
\bibitem[Yi et al.(2022)]{2022A&A...663A.155Y} Yi, S.-X., Nelemans, G., Brinkerink, C., et al.\ 2022, \aap, 663, A155. doi:10.1051/0004-6361/202141634
\bibitem[Zhang et al.(2022)]{2022ApJ...939L..25Z} Zhang, Z., Yi, S.-X., Zhang, S.-N., et al.\ 2022, \apjl, 939, L25. doi:10.3847/2041-8213/ac9b55

\end{thebibliography}
\end{document}